\documentclass{article}

\usepackage{microtype}
\usepackage{graphicx}
\usepackage{booktabs} 

\usepackage{hyperref}
\usepackage{glossaries}
\usepackage{listings}
\usepackage[position=b]{subcaption}
\usepackage[nameinlink]{cleveref}
\usepackage{siunitx}
\usepackage{tabularx}
\usepackage{amsmath}
\usepackage{tikz}
\usepackage{enumitem}
\usepackage{orcidlink}

\usepackage[T1]{fontenc}
\usepackage{tgbonum}

\DeclareSIUnit{\ge}{GE}
\newcommand{\tblmark}[1]{\,%
  \tikz[baseline={([yshift=-0.6ex]c.center)}]{\node[circle,draw,fill=black,text=white,inner sep=1pt,font=\scriptsize\bfseries,line width=0.4pt](c){#1};}%
}

\glsdisablehyper

\setlist[itemize]{itemsep=1em, topsep=2pt, parsep=0pt, partopsep=0pt}

\ifdefined\blindreview
    \usepackage{mlsys2026}
\else
    \usepackage[accepted]{mlsys2026}
\fi

\mlsystitlerunning{A Collective-Capable NoC for Large-Scale ML Accelerators}

\makeatletter
\newcommand{\insertnewlines}[1]{%
  \noindent\mbox{}%
  \@tempcnta=#1\relax
  \loop\ifnum\@tempcnta>0
    \\
    \advance\@tempcnta\m@ne
  \repeat
}
\makeatother

\newcommand\variable[1]{\mathop{\mathit{#1}}\nolimits}

\newacronym[plural=WANs, firstplural={Wide Area Networks (WANs)}]{wan}{WAN}{Wide Area Network}
\newacronym[plural=WSNs, firstplural={Wireless Sensor Networks (WSNs)}]{wsn}{WSN}{Wireless Sensor Network}
\newacronym{simd}{SIMD}{Single Instruction Multiple Data}
\newacronym{os}{OS}{Operating System}
\newacronym{ble}{BLE}{Bluetooth Low-Energy}
\newacronym{wifi}{Wi-FI}{Wireless Fidelity}
\newacronym[plural=DVS, firstplural={Dynamic Vision Sensors (DVS)}]{dvs}{DVS}{Dynamic Vision Sensor}
\newacronym{ptz}{PTZ}{Pan-Tilt Unit}

\newacronym[plural=FLLs,firstplural=Frequency Locked Loops (FLLs)]{fll}{FLL}{Frequency Locked Loop}
\newacronym{dram}{DRAM}{Dynamic Random Access Memory}
\newacronym{fpu}{FPU}{Floating Point Unit}
\newacronym{fpss}{FPSS}{Floating Point Subsystem}
\newacronym{frep}{FREP}{Floating Point Repetition}
\newacronym{dma}{DMA}{Direct Memory Access}
\newacronym{dca}{DCA}{Direct Compute Access}
\newacronym{ssr}{SSR}{Stream Semantic Register}
\newacronym{issr}{ISSR}{Indirection Stream Semantic Register}
\newacronym[plural=LUTs, firstplural={Lookup Tables (LUTs)}]{lut}{LUT}{Lookup Table}
\newacronym[plural=FPGAs, firstplural={Field Programmable Gate Arrays (FPGAs)}]{fpga}{FPGA}{Field Programmable Gate Array}
\newacronym{dsp}{DSP}{Digital Signal Processing}
\newacronym{mcu}{MCU}{Microcontroller Unit}
\newacronym{spi}{SPI}{Serial Peripheral Interface}
\newacronym{cpi}{CPI}{Camera Parallel Interface}
\newacronym{rf}{RF}{Register File}
\newacronym{fifo}{FIFO}{First-In First-Out Queue}
\newacronym{uart}{UART}{Universal Asynchronous Receiver-Transmitter}
\newacronym{raw}{RAW}{Read After Write}
\newacronym[plural=ISAs, firstplural={Instruction Set Architectures (ISAs)}]{isa}{ISA}{Instruction Set Architecture}
\newacronym{xbar}{XBAR}{crossbar}
\newacronym[firstplural=Scratch-Pad Memories (SPMs)]{spm}{SPM}{Scratch-Pad Memory}
\newacronym{ppa}{PPA}{Power Performance Area}
\newacronym{ipi}{IPI}{Inter-Processor Interrupt}
\newacronym[firstplural=Software-Generated Interrupts (SGIs)]{sgi}{SGI}{Software-Generated Interrupt}
\newacronym{pe}{PE}{Processing Element}
\newacronym{tcdm}{TCDM}{Tightly-Coupled Data Memory}
\newacronym{lsu}{LSU}{Load-Store Unit}
\newacronym{icache}{I\$}{Instruction Cache}
\newacronym{dcache}{D\$}{Data Cache}
\newacronym{wfi}{WFI}{Wait For Interrupt}
\newacronym{gpc}{GPC}{GPU Processing Cluster}
\newacronym{cpu}{CPU}{Central Processing Unit}
\newacronym{gpu}{GPU}{Graphics Processing Unit}
\newacronym{llc}{LLC}{Last-Level Cache}
\newacronym{sm}{SM}{Streaming Multiprocessor}
\newacronym[firstplural=Networks on Chip (NoCs)]{noc}{NoC}{Network on Chip}
\newacronym{ni}{NI}{Network Interface}
\newacronym[firstplural=Gate Equivalents (GEs)]{ge}{GE}{Gate Equivalent}
\newacronym{vcd}{VCD}{Value Change Dump}
\newacronym{rtl}{RTL}{Register Transfer Level}

\newacronym{dfg}{DFG}{Data Flow Graph}
\newacronym{lcg}{LCG}{Linear Congruential Generator}
\newacronym{prn}{PRN}{Pseudo-Random Number}
\newacronym{gemm}{GEMM}{General Matrix Multiplication}

\newacronym{ste}{STE}{Straight-Through-Estimator}

\newacronym[plural=PTUs, firstplural={Pan-Tilt Units}]{ptu}{PTU}{Pan-Tilt Unit}
\newacronym{mdf}{MDF}{Medium-density fibreboard}
\newacronym{cvat}{CVAT}{Computer Vision Annotation Tool}
\newacronym{coco}{COCO}{Common Objects in Context}
\newacronym{soa}{SoA}{State-of-the-Art}
\newacronym{sf}{SF}{Sensor Fusion}

\newacronym{dl}{DL}{Deep Learning}
\newacronym{bn}{BN}{Batch Normalization}
\newacronym{FGSM}{FBK}{Fast Gradient Sign Method}
\newacronym{lr}{LR}{Learning Rate}
\newacronym{sgd}{SGD}{Stochastic Gradient Descent}
\newacronym{gd}{GD}{Gradient Descent}
\newacronym{llm}{LLM}{Large Language Model}

\newacronym{sta}{STA}{Static Timing Analysis}

\newacronym[plural=GPIOs, firstplural={General Purpose Inupt Outputs (GPIOs)}]{gpio}{GPIO}{General Purpose Input Output}
\newacronym[plural=LDOs, firstplural={Low Dropout Regulators (LDOs)}]{ldo}{LDO}{Low Dropout Regulator}

\newacronym{inq}{INQ}{Incremental Network Quantization}

\newacronym{CV}{CV}{Computer Vision}
\newacronym{EoT}{EoT}{Expectation over Transformation}
\newacronym{RPN}{RPN}{Region Proposal Network}
\newacronym{TV}{TV}{Total Variation}
\newacronym{NPS}{NPS}{Non-Printability Score}
\newacronym{STN}{STN}{Spatial Transformer Network}
\newacronym{MTCNN}{MTCNN}{Multi-Task Convolutional Neural Network}
\newacronym{YOLO}{YOLO}{You Only Look Once}
\newacronym{SSD}{SSD}{Single Shot Detector}
\newacronym{SOTA}{SOTA}{State of the Art}
\newacronym{NMS}{NMS}{Non-Maximum Suppression}
\newacronym{ic}{IC}{Integrated Circuit}
\newacronym{tcxo}{TCXO}{Temperature Controlled Crystal Oscillator}
\newacronym{jtag}{JTAG}{Joint Test Action Group industry standard}
\newacronym{swd}{SWD}{Serial Wire Debug}
\newacronym{sdio}{SDIO}{Serial Data Input Output}

\newacronym[plural=PCBs, firstplural={Printed Circuit Boards (PCB)}]{pcb}{PCB}{Printed Circuit Board}
\newacronym[plural=ASICs, firstplural={Application Specific Integrated Circuits}]{asic}{ASIC}{Application Specific Integrated Circuit}

\newacronym[plural=BNNs, firstplural={Binary Neural Networks (BNNs)}]{bnn}{BNN}{Binary Neural Network}
\newacronym[plural=NNs, firstplural={Neural Networks}]{nn}{NN}{Neural Network (NNs)}
\newacronym[plural=SCMs, firstplural={Standard Cell Memories (SCMs)}]{scm}{SCM}{Standard Cell Memory}
\newacronym{ann}{ANN}{Artificial Neural Networks}
\newacronym{ml}{ML}{Machine Learning}
\newacronym{ai}{AI}{Artificial Intelligence}
\newacronym{iot}{IoT}{Internet of Things}
\newacronym{fft}{FFT}{Fast Fourier Transform}
\newacronym[plural=OCUs, firstplural={Output Channel Compute Units (OCUs)}]{ocu}{OCU}{Output Channel Compute Unit}
\newacronym{alu}{ALU}{Arithmetic Logic Unit}
\newacronym{mac}{MAC}{Multiply-Accumulate}
\newacronym[firstplural={systems-on-chip (SoCs)}]{soc}{SoC}{system-on-chip}
\newacronym[firstplural={multi-processor systems-on-chip (MPSoCs)}]{mpsoc}{MPSoC}{multi-processor system-on-chip}

\newacronym{PGD}{PGD}{Projected Gradient Descend}
\newacronym{CW}{CW}{Carlini-Wagner}
\newacronym{OD}{OD}{Object Detection}

\newacronym{rrf}{RRF}{RADAR Repetition Frequency}
\newacronym{nlp}{NLP}{Natural Language Processing}
\newacronym{qam}{QAM}{Quadrature Amplitude Modulation}
\newacronym{rri}{RRI}{RADAR Repetition Interval}
\newacronym{radar}{RADAR}{Radio Detection and Ranging}
\newacronym{loocv}{LOOCV}{Leave-one-out cross validation}

\newacronym{bsp}{BSP}{Board Support Package}
\newacronym{ttn}{TTN}{The Things Network}
\newacronym{wip}{WIP}{Work in Progress}
\newacronym{json}{JSON}{JavaScript Object Notation}
\newacronym{qat}{QAT}{Quantization-Aware Training}

\newacronym{cls}{CLS}{Classification Error}
\newacronym{loc}{LOC}{Localization Error}
\newacronym{bkgd}{BKGD}{Background Error}
\newacronym{roc}{ROC}{Receiver Operating Characteristic}
\newacronym{frr}{FRR}{False Rejection Rate}
\newacronym{eer}{EER}{Equal Error Rate}
\newacronym{snr}{SNR}{Signal-to-Noise Ratio}
\newacronym{flop}{FLOP}{Floating-Point Operation}
\newacronym{fp}{FP}{Floating-Point}
\newacronym{fps}{FPS}{Frames Per Second}
\newacronym{oi}{OI}{Operational Intensity}
\newacronym{ipc}{IPC}{Instructions per Cycle}

\newacronym{gsc}{GSC}{Google Speech Commands}
\newacronym{mswc}{MSWC}{Multilingual Spoken Words Corpus}
\newacronym{demand}{DEMAND}{Diverse Environments Multichannel Acoustic Noise Database}

\newacronym[plural=SNNs, firstplural={Spiking Neural Networks (SNNs)}]{snn}{SNN}{Spiking Neural Network}
\newacronym[plural=DNNs, firstplural={Deep Neural Networks (DNNs)}]{dnn}{DNN}{Deep Neural Network}
\newacronym[plural=TCNs,firstplural=Temporal Convolutional Networks]{tcn}{TCN}{Temporal Convolutional Network}
\newacronym[plural=CNNs,firstplural=Convolutional Neural Networks (CNNs)]{cnn}{CNN}{Convolutional Neural Network}
\newacronym[plural=TNNs,firstplural=Ternarized Neural Networks]{tnn}{TNN}{Ternarized Neural Network}
\newacronym{ds-cnn}{DS-CNN}{Depthwise Separable Convolutional Neural Network}
\newacronym{rnn}{RNN}{Recurrent Neural Network}
\newacronym{gcn}{GCN}{Graph Convolutional Network}
\newacronym{mha}{MHA}{Multi-Head Attention}
\newacronym{crnn}{CRNN}{Convolutional Recurrent Neural Network}
\newacronym{clca}{CLCA}{Convolutional Linear Cross-Attention}

\newacronym{bf}{BF}{Beamforming}
\newacronym{anc}{ANC}{Active Noise Cancellation}
\newacronym{agc}{AGC}{Automatic Gain Control}
\newacronym{se}{SE}{Speech Enhancement}
\newacronym{mct}{MCT}{Multi-Condition Training}
\newacronym{mcta}{MCTA}{Multi-Condition Training \& Adaptation}
\newacronym{pcen}{PCEN}{Per-Channel Energy Normalization}
\newacronym{mfcc}{MFCC}{Mel-Frequency Cepstral Coefficient}
\newacronym{asr}{ASR}{Automated Speech Recognition}
\newacronym{kws}{KWS}{Keyword Spotting}
\newacronym{odl}{ODL}{On-Device Learning}

\newacronym{nl-kws}{NL-KWS}{Noiseless Keyword Spotting}
\newacronym{na-kws}{NA-KWS}{Noise-Aware Keyword Spotting}
\newacronym{odda}{ODDA}{On-Device Domain Adaptation}
\newacronym{hpm}{HPM}{High-Performance Mode}
\newacronym{lpm}{LPM}{Low-Power Mode}

\newcommand{\ResultSlopeHardwareBarrier}{1.3}
\newcommand{\ResultSlopeSoftwareBarrier}{3.3}
\newcommand{\ResultMinHwMcastSpeedup}{2.3}
\newcommand{\ResultMaxHwMcastSpeedup}{3.2}

\newcommand{\ResultMinMcastSizeKibiBytes}{1}
\newcommand{\ResultMaxMcastSizeKibiBytes}{32}

\newcommand{\ResultGeomeanHwMcastTwoDSpeedup}{5.3}
\newcommand{\ResultMinHwReductionSpeedup}{2.0}
\newcommand{\ResultMaxHwReductionSpeedup}{3.0}

\newcommand{\ResultMinReductionSizeKibiBytes}{1}
\newcommand{\ResultMaxReductionSizeKibiBytes}{32}
\newcommand{\ResultOneDimToTwoDimSlowdown}{1.9}
\newcommand{\ResultOneDimToTwoDimSlowdownKibiBytes}{32}

\newcommand{\ResultGeomeanHwReductionTwoDSpeedup}{2.8}
\newcommand{\ResultMedianGemmUtilizationPercentage}{98.1}
\newcommand{\ResultMinGemmHwMcastSpeedup}{1.1}
\newcommand{\ResultMaxGemmHwMcastSpeedup}{3.8}
\newcommand{\ResultMaxGemmHwReductionSpeedup}{2.4}
\newcommand{\ResultSummaMaxEnergySaving}{1.17}
\newcommand{\ResultFclMaxEnergySaving}{1.13}
\newcommand{\ResultEnDmaLd}{2.2}
\newcommand{\ResultEnDmaSt}{2.4}
\newcommand{\ResultEnLinkToLink}{1.1}
\newcommand{\ResultEnWriteTcdm}{1.8}
\newcommand{\ResultEnGemm}{24.6}
\newcommand{\ResultEnSwRed}{22.4}
\newcommand{\ResultEnHwRed}{19.0}
\newcommand{\ResultSummaSwDmaLd}{66}
\newcommand{\ResultSummaSwDmaSt}{983}
\newcommand{\ResultSummaSwLinkToLink}{1114}
\newcommand{\ResultSummaSwTcdmWrite}{983}
\newcommand{\ResultSummaSwGemm}{1049}
\newcommand{\ResultSummaSwSwRed}{0}
\newcommand{\ResultSummaSwHwRed}{0}
\newcommand{\ResultSummaHwDmaLd}{66}
\newcommand{\ResultSummaHwDmaSt}{66}
\newcommand{\ResultSummaHwLinkToLink}{983}
\newcommand{\ResultSummaHwTcdmWrite}{983}
\newcommand{\ResultSummaHwGemm}{1049}
\newcommand{\ResultSummaHwSwRed}{0}
\newcommand{\ResultSummaHwHwRed}{0}
\newcommand{\ResultFclSwDmaLd}{524}
\newcommand{\ResultFclSwDmaSt}{524}
\newcommand{\ResultFclSwLinkToLink}{4524}
\newcommand{\ResultFclSwTcdmWrite}{522}
\newcommand{\ResultFclSwGemm}{1049}
\newcommand{\ResultFclSwSwRed}{65}
\newcommand{\ResultFclSwHwRed}{0}
\newcommand{\ResultFclHwDmaLd}{524}
\newcommand{\ResultFclHwDmaSt}{72}
\newcommand{\ResultFclHwLinkToLink}{3932}
\newcommand{\ResultFclHwTcdmWrite}{35}
\newcommand{\ResultFclHwGemm}{1049}
\newcommand{\ResultFclHwSwRed}{0}
\newcommand{\ResultFclHwHwRed}{65}

\newcommand{\ResultMulticastForkingOverhead}{6.4}
\newcommand{\ResultMulticastCollectBOverhead}{36.4}
\newcommand{\ResultMulticastRouterOverhead}{5.8}
\newcommand{\ResultParallelRedRouterOverhead}{2.7}
\newcommand{\ResultRedArbiterGe}{1.13}
\newcommand{\ResultWideSimpleControllerGe}{13.62}
\newcommand{\ResultWideSimpleCrtlComb}{56.3}
\newcommand{\ResultWideSimpleCrtlSeq}{43.7}
\newcommand{\ResultAreaOverheadTot}{16.9}

\glsunset{cpu}

\begin{document}

\twocolumn[
\mlsystitle{A Lightweight High-Throughput Collective-Capable NoC\\for Large-Scale ML Accelerators}


\mlsyssetsymbol{equal}{*}

\begin{mlsysauthorlist}
\mlsysauthor{Luca Colagrande\,\orcidlink{0000-0002-7986-1975}}{equal,iis}
\mlsysauthor{Lorenzo Leone\,\orcidlink{0009-0000-3976-847X}}{equal,iis}
\mlsysauthor{Chen Wu\,\orcidlink{0009-0006-5417-2870}}{iis}
\mlsysauthor{Tim Fischer\,\orcidlink{0009-0007-9700-1286}}{iis}
\mlsysauthor{Raphael Roth\,\orcidlink{0009-0000-8445-1674}}{itet}
\mlsysauthor{Luca Benini\,\orcidlink{0000-0001-8068-3806}}{iis}
\end{mlsysauthorlist}

\mlsysaffiliation{iis}{Integrated Systems Laboratory (IIS), ETH Zurich, Zurich, Switzerland}
\mlsysaffiliation{itet}{D-ITET, ETH Zurich, Zurich, Switzerland}

\mlsyscorrespondingauthor{Luca Colagrande}{colluca@iis.ee.ethz.ch}
\mlsyscorrespondingauthor{Lorenzo Leone}{lleone@iis.ee.ethz.ch}

\mlsyskeywords{network-on-chip, collective communication, multicast, reduction, direct compute access, in-network compute, von neumann bottleneck, memory wall}

\vskip 0.3in

\begin{abstract}
    The exponential increase in \gls{ml} model size and complexity has driven unprecedented demand for high-performance acceleration systems.
    As technology scaling enables the integration of thousands of computing elements onto a single die, the boundary between distributed and on-chip systems has blurred, making efficient on-chip collective communication increasingly critical.
    In this work, we present a lightweight, collective-capable \gls{noc} that supports efficient barrier synchronization alongside scalable, high-bandwidth multicast and reduction operations, co-designed for the next generation of \gls{ml} accelerators.
    We introduce \gls{dca}, a novel paradigm that grants the interconnect fabric direct access to the cores' computational resources, enabling high-throughput in-network reductions with a small \ResultAreaOverheadTot\% router area overhead.
    Through in-network hardware acceleration, we achieve \ResultGeomeanHwMcastTwoDSpeedup$\times$ and \ResultGeomeanHwReductionTwoDSpeedup$\times$ geomean speedups on multicast and reduction operations involving between \SI{1}{} and \SI{32}{\kibi\byte} of data, respectively.
    Furthermore, by keeping communication off the critical path in \acrshort{gemm} workloads, these features allow our architecture to scale efficiently to large meshes, resulting in up to \ResultMaxGemmHwMcastSpeedup$\times$ and \ResultMaxGemmHwReductionSpeedup$\times$ estimated performance gains through multicast and reduction support, respectively, compared to a baseline unicast NoC architecture, and up to \ResultSummaMaxEnergySaving$\times$ estimated energy savings.
\end{abstract}
]



\printAffiliationsAndNotice{\mlsysEqualContribution} 

\setlength{\abovedisplayskip}{5pt plus 1pt minus 1pt}
\setlength{\belowdisplayskip}{5pt plus 1pt minus 1pt}
\setlength{\abovedisplayshortskip}{4pt plus 2pt minus 2pt}
\setlength{\belowdisplayshortskip}{4pt plus 2pt minus 2pt}

\setlength{\jot}{2pt} 

\section{Introduction}
\label{sec:introduction}
With the explosion of transformer-based models and the exponential increase in model size and complexity \cite{ai-model-size}, the demand for high-performance systems has surged dramatically.
To meet these requirements, massively parallel accelerators must evolve rapidly, pushing the limits of computational capability.
Modern architectures are integrating an ever-growing number of \glspl{pe} to boost peak performance.
For example, NVIDIA's latest Blackwell \acrshort{gpu} roughly doubles the number of CUDA cores over the previous generation \cite{blackwell}.

However, this rapid growth in computational power has not been matched by proportional improvements in data movement.
Performance has far outpaced memory and communication bandwidth: over the past two decades, peak FLOPS throughput improved by about $60000\times$, whereas DRAM bandwidth grew only around $100\times$ \cite{mem-wall}. This widening gap between computation and data movement has become a significant bottleneck for modern systems, often leading to memory-bound or communication-bound workloads ~\cite{ai-data-bottleneck, mutlu2023}.

Among the various parallel computing paradigms, collective communication plays a crucial role in enabling efficient data exchange among \glspl{pe}.
Recent analyses of MPI usage~\cite{mpi-usage} show that the most frequently used collective operations, such as \textit{reduction}, \textit{barrier}, and \textit{broadcast}, are essential for synchronizing data across multiple nodes.
To mitigate the overhead of these operations in large-scale distributed systems, dedicated hardware engines~\cite{nvidia-sharp, ibm-blugene} and optimized software libraries~\cite{nvidia-nccl} for multi-\acrshort{gpu} and multi-node systems have been developed.

\begin{figure*}[t!]
    \centering
    \includegraphics[width=1\textwidth]{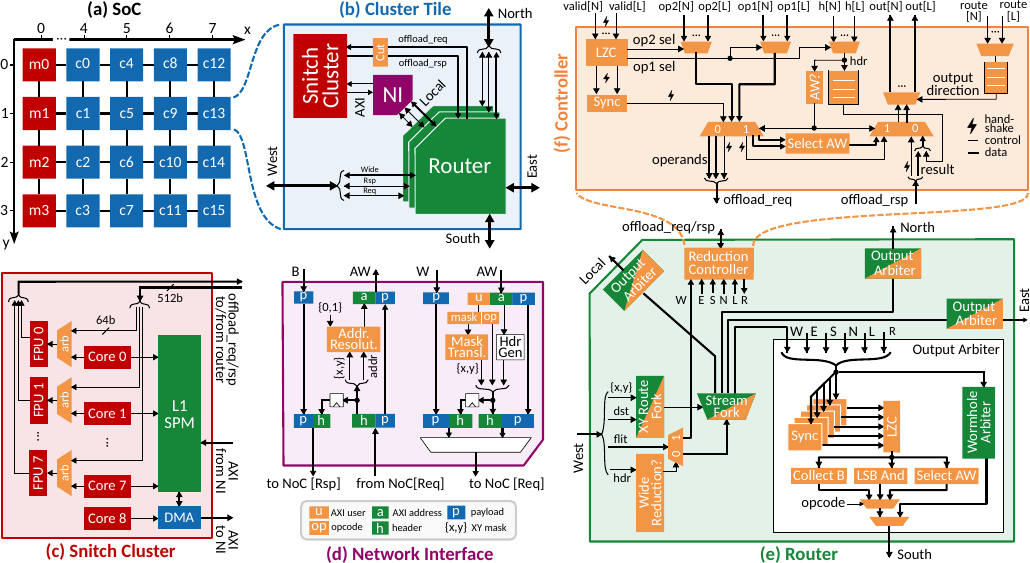}

    \begingroup
        \phantomsubcaption
        \label{fig:soc}
        \phantomsubcaption
        \label{fig:cluster-tile}
        \phantomsubcaption
        \label{fig:cluster}
        \phantomsubcaption
        \label{fig:ni}
        \phantomsubcaption
        \label{fig:router}
        \phantomsubcaption
        \label{fig:reduction-controller}
    \endgroup

    \caption{(a) Overview of the $5\times4$ collective-capable \gls{noc} system.
             (b) Cluster tile and its main components: (c) compute cluster, (d) network interface and (e) router with collective extensions.
             (f) Centralized reduction controller enabling arithmetic in-network computation.
             Highlighted in orange are all modules affected (partially highlighted) or introduced (fully highlighted) by our extensions.}
    \label{fig:architecture}
\end{figure*}

However, as technology scaling has allowed the integration of thousands of \glspl{pe} and specialized compute coprocessors (e.g. tensor units) onto a single die, the boundary between distributed and on-chip systems has blurred.
Tile-based manycore \glspl{soc}, as the one represented in \Cref{fig:soc}, have effectively become self-contained parallel systems, making efficient on-chip collective communication increasingly critical.
Without proper support, collective operations can quickly saturate memory and interconnect resources, limiting performance scalability.

As shown in \Cref{sec:gemm-results}, we observe that on large-scale accelerator configurations GEMM kernels become memory-bound, resulting in \textless\,50\% utilization on a 256x256 mesh.
By accelerating collective communication primitives we can reduce the communication time, resulting in up to \ResultMaxGemmHwMcastSpeedup$\times$ speedups on the overall kernel runtime.
Similarly, FlatAttention~\cite{zhang2025} has demonstrated the potential of coordinated on-chip collective operations to reduce external memory traffic and improve utilization of on-device tensor engines, reporting up to $4\times$ speedups over FlashAttention-3.

In this work, we present a complete design of the first lightweight collective-capable \gls{noc} tailored for general-purpose manycore \gls{ml} systems.
To the best of our knowledge, this is also the first work to demonstrate that high-throughput arithmetic reductions can be efficiently implemented on-chip, by sharing resources between the interconnect fabric and accelerator clusters.
Specifically:
\begin{itemize}
    \item We present the design and implementation%
\footnote{Our implementation is fully open source and can be found at:
\ifdefined\blindreview
    https://hidden-for-double-blind-review.com
\else
    \url{https://github.com/pulp-platform/FlooNoC/releases/tag/v0.8.0}
    \url{https://github.com/pulp-platform/picobello/commit/cacdc3ad4a3a638e68d03090781dbbc2450c73ed}
\fi
}%
, of a general-purpose collective-capable \gls{noc} that enables high-performance and efficient in-network computing primitives.
    In particular, our NoC supports multicast and reduction operations for both low-latency synchronization and high-bandwidth communication tasks, targeting general-purpose many-core \gls{ml} accelerators.
    \item We demonstrate the flexibility of our design by integrating it into a multi-cluster \gls{soc} for \gls{noc} evaluation \cite{fischer2025}.
    We introduce a novel paradigm, \acrfull{dca}, granting the interconnect fabric direct access to the cores' compute resources, to enable low-cost high-throughput in-network compute.
    \item We implement our collective-capable \gls{noc} in an advanced technology node, demonstrating that our design incurs a small router area overhead of \ResultAreaOverheadTot\% compared to the \gls{soa} FlooNoC baseline, accounting for a negligible \textless\,1\% area increase on a full compute tile, without degrading timing performance.
    \item We provide a comprehensive performance evaluation of the proposed architecture through cycle-accurate simulation and analytical modeling, comparing to highly optimized software implementations of multicast and reduction primitives.
    In a 4$\times$4 mesh, we measure geomean speedups of \ResultGeomeanHwMcastTwoDSpeedup$\times$ and \ResultGeomeanHwReductionTwoDSpeedup$\times$ respectively, on transfers involving between \SI{\ResultMinMcastSizeKibiBytes}{} and \SI{\ResultMaxMcastSizeKibiBytes}{\kibi\byte} of data.
\end{itemize}

\section{Background}
\label{sec:background}

\subsection{FlooNoC}
This work builds on FlooNoC~\cite{fischer2025}, a \gls{soa}, AXI-compliant%
\footnote{See \Cref{sec:hw-terminology} for a background on hardware terminology.}%
~\cite{axi}, open-source \gls{noc} designed for the requirements of next generation \gls{ml} accelerators, in which bulk data transfers coexist with short latency-critical messages.
This is achieved by providing two dedicated networks for the different traffic classes: a ``wide'' network for high-bandwidth bursted data transfers and a ``narrow'' network for latency-critical messages, respectively 512- and 64-bit wide in our reference implementation.

A \gls{ni} at every endpoint maps the AXI channels from these two networks to three distinct physical links at the \gls{noc} level:
a \texttt{wide} link carrying wide read and write transfers, a \texttt{req} link carrying both wide and narrow requests and narrow write transfers, and a \texttt{rsp} channel carrying both wide and narrow responses and narrow read transfers.
A multi-link router at every endpoint, composed of separate routers for each of the three physical links, routes flits (in FlooNoC terminology equivalent to packets) between the respective \gls{ni} and neighbouring endpoints.

\subsection{Baseline \gls{soc} Architecture}
\label{sec:baseline-soc}

For our evaluation, we adopt a similar system to FlooNoC \cite{fischer2025}, integrating multiple compute and memory tiles in a 2D mesh topology, as shown in \Cref{fig:soc}.
Each memory tile includes a \SI{1}{\mega\byte} L2 \gls{spm}, while each compute tile hosts a Snitch cluster \cite{zaruba2021} composed of 8 energy-efficient RV32I Snitch cores. Each core is paired with a 64-bit SIMD-capable \gls{fpu}, and shares a \SI{128}{\kibi\byte} L1 \gls{spm} within the cluster.
Each cluster also features a ninth Snitch core with a \gls{dma} engine to orchestrate high-bandwidth data transfers between tiles.

The \gls{dma} in the \textit{initiator} tile transfers data in \textit{bursts} from a \textit{source} tile to a \textit{destination} tile over the wide network.
A \gls{dma} transaction proceeds as follows: 1) the initiator issues a read request (AR) to the source, 2) the corresponding read \textit{beats} (R) are returned to the initiator, 3) which are forwarded as write beats (AW/W) to the destination, and 4) a final response (B) is sent back to the initiator.
The trasfer time can be modeled as $T_{transfer} = \alpha + n \beta$, where $\alpha$ denotes the round-trip latency of the transfer (including the initiator-source-initiator and initiator-destination-initiator paths), $n$ is the number of beats in the transfer and $\beta$ represents the inverse bandwidth, expressed in cycles per beat.

\subsection{Multi-address encoding}
\label{sec:encoding}
A key challenge in enabling collective operations lies in efficiently encoding multiple destination addresses.
Previous work \cite{colagrande2025} on a multicast-capable \gls{xbar} extended the AXI protocol to support multi-address representation by pairing a destination address with a mask.
The mask, equal in width to the destination address, is carried in the user-defined \texttt{AWUSER} AXI signal~\cite{axi}.
When a mask bit is set to 1, the corresponding address bit is treated as a ``don't care'' (X), representing both logic 0 and 1.
By masking $n$ bits in the address, up to $2^n$ distinct destinations can be represented within a single transaction.

Although this encoding scheme restricts the set of address combinations that can be represented, it remains highly scalable\footnote{Arbitrary destination sets can still be represented through multiple multi-address transactions, at the cost of increased overhead.}:
the encoding grows logarithmically with the address space size and is independent of the number of destinations.
This property makes it particularly well suited for large-scale architectures such as massively parallel accelerators.

\section{Architecture}
\label{sec:implementation}

\subsection{Collective-Capable \gls{noc}}
\label{sec:impl-noc}

We extend the FlooNoC \cite{fischer2025} architecture to support collective communication operations, with a focus on multicast and reduction.
All extensions are designed to remain fully compliant with the AXI4 protocol \cite{axi} supported by FlooNoC.
As detailed in \Cref{sec:encoding}, we extend the \texttt{AWUSER} field to carry the multi-address mask, and an opcode specifying the collective operation to perform.

Furthermore, the AXI4 protocol inherently couples multicast and reduction operations.
When a manager issues a multicast request (AW), it is delivered to multiple destinations, each producing a corresponding response (B).
Since the manager expects a single response to its request, the network must aggregate the responses from all destinations, effectively performing a reduction on the responses within the network.
Conversely, when a reduction operation is initiated by multiple managers, the resulting response from the destination must be multicast back to all initiators.


\subsubsection{Network Interface}
\label{sec:network-interface}

The \gls{ni} forms the bridging point between the \gls{noc} and the endpoint protocol, translating incoming AXI packets into the \gls{noc} protocol and vice versa.
\Cref{fig:ni} illustrates a high-level block diagram of the \gls{ni}, where only the channels and logic affected by the collective extensions are highlighted.

While the multicast-capable AXI \gls{xbar} described in \cite{colagrande2025} routes packets by directly comparing their destination address and mask, flits in the \gls{noc} are routed using the X and Y coordinates of their destination and source nodes.
Thus, to represent the multiple destination nodes of one-to-many operations (e.g., multicast) and the multiple source nodes of many-to-one operations (e.g., reduction), the \gls{noc} protocol must be provided with equivalent coordinate masks.
As shown in \Cref{fig:ni}, the \gls{ni} translates the address mask in the \texttt{AWUSER} field into the corresponding X and Y masks, which are appended to the AW flit header.
As W and AW beats form part of the same AXI transaction, they must use the same X and Y masks; this information is therefore stored in a register and reused when injecting the subsequent W flits.

Conversely, when a collective request arrives at the \gls{ni} from the \gls{noc}, the local endpoint coordinates (e.g. \{0,1\}) are used to ``resolve'' the incoming multi-address, translating it back into the endpoint's local address space.
An additional buffer in the \gls{ni} stores the incoming mask information to generate the appropriate collective response, namely a multicast response to reduction requests and vice versa.

\subsubsection{Multicast Router Extension}
\label{sec:impl-multicast}

\Cref{fig:router} illustrates a simplified schematic of the router, highlighting the components affected by our extensions.
The \lstinline{xy_route_fork} at each input port is responsible for calculating the output port onto which an incoming packet should be routed, based on its destination coordinates (\lstinline{dst}).
Since multicast involves forking incoming packets to multiple output directions, we extend the \lstinline{xy_route_fork} to select multiple output ports according to the X and Y masks in the request.
As discussed in \Cref{sec:network-interface}, these masks are generated by the \gls{ni} and, together with the destination coordinates in the flit header, are used to represent the multiple destination nodes of a multicast request.
Following the method presented in \cite{colagrande2025}, if a bit in the X (or Y) mask is set to 1, the corresponding bit in \lstinline{dst.X} (or \lstinline{dst.Y}) is treated as a ``don't care'', encoding both logic 0 and 1.
Masking $n$ bits in the \lstinline{dst} coordinates therefore allows the \lstinline{(dst, mask)} pair to represent $2^n$ destination nodes.
Finally, the \lstinline{xy_route_fork} drives the downstream \lstinline{stream_fork} module, which demultiplexes the incoming packet to the output ports specified by the \lstinline{select} signal, while ensuring that the input is accepted only once all output ports are ready to receive it.

\subsubsection{Parallel Reduction Router Extension}
\label{sec:impl-parallel-reduction}

Because of the coupling between multicast and reduction operations inherent to the AXI protocol, introduced in \Cref{sec:impl-noc}, supporting multicast operations requires minimal support for reductions as well.
To reduce incoming packets from multiple input directions, every output port is equipped with an \lstinline{output_arbiter}, which extends the functionality of the original \lstinline{wormhole_arbiter} with reduction capabilities.
As shown in \Cref{fig:router}, the output arbiter directs unicast packets from all input directions to the \lstinline{wormhole_arbiter}, where these are arbitrated and forwarded to the output port one at a time.
Reduction packets, on the other hand, are redirected to a dedicated \lstinline{reduction_arbiter} where packets from selected input directions are reduced together.

The input directions involved in a reduction operation are determined by the X and Y mask encoded in the flit header, together with the coordinates of the source node that issued the packet.
This logic is implemented in the synchronization module, which takes one input as reference for said calculation, and waits for the corresponding reduction flits expected at the other inputs to arrive.
These are forwarded downstream only once the packets from all the selected input directions are available.
A synchronization module per input port is instantiated and a \lstinline{leading_zero_counter} is then used to arbitrate between concurrent reduction operations.
This design allows multiple reductions to coexist without deadlocks, even when their paths cross within the network, as the replicated synchronization modules ensure that each incoming reduction is only arbitrated if it is guaranteed it can be carried out to completion.

Once an incoming reduction is selected, a dedicated computation block performs the reduction operation on all inputs in parallel.
Multiple blocks can be instantiated to support a variety of operations, and selected through the reduction opcode in the header flit.
In this work, we implement three lightweight operations:
1) \lstinline{CollectB}, reduces the multiple B responses to a multicast operation;
2) \lstinline{LsbAnd}, performs a bitwise AND-reduction on the least significant bits in the incoming packets; and
3) \lstinline{SelectAW}, aggregates the multiple AW requests from a reduction operation.
While the \lstinline{CollectB} and \lstinline{SelectAW} operations are respectively required to support multicast and reductions in AXI, we use the \lstinline{LsbAnd} operation to implement efficient barrier synchronization primitives, as shown in \Cref{sec:narrow-reduction-results}.

\subsubsection{Wide Reduction Router Extension}
\label{sec:impl-wide-reduction}

For more complex reduction operations, e.g. involving floating-point operands on the wide network, implementing a 5-input reduction tree in hardware may be prohibitively expensive.
Despite this limitation, supporting reduction operations limited to two inputs per router can still be beneficial, as shown in \Cref{sec:wide-reduction-results}.
These operations often rely on pipelined arithmetic units, requiring additional logic to handle the reduction.
Unlike the lightweight parallel reduction logic replicated at every output port, we provide a single centralized instance of the wide reduction logic, shared across all outputs, as illustrated in \Cref{fig:router,fig:reduction-controller}.

A synchronization module again ensures that all input operands, up to two in this case, are received before being forwarded to the arithmetic unit.
Unlike the parallel reduction, we instantiate a single synchronization module, with an upstream arbiter (\lstinline{lzc}) selecting the reference input.
At any given time, we support a single reduction operation of this kind per router, preventing deadlocks while simplifying the design compared to the parallel reduction scheme.

To accommodate pipelined functional units, a \lstinline{hdr} buffer is required to temporarily store the header flit until the result is produced by the arithmetic unit.
The result is then concatenated with the header and routed as a regular unicast packet to the output wormhole arbiter.
By increasing the depth of the buffer, multiple reduction operations can be issued back-to-back, effectively hiding the pipeline latency and achieving a throughput of one reduction per cycle.
This is particularly beneficial for bursted reduction operations, as occur on the wide network.
To avoid stalls, the buffer depth must be greater than the functional unit's pipeline depth.

Finally, we provide an offload port for 2-input reductions that can be executed by external compute resources, as described in \Cref{sec:dca}.

\subsection{System-Level Integration}
\label{sec:impl-system}

\Cref{fig:cluster,fig:cluster-tile} illustrate the integration of the collective-capable \gls{noc} within the baseline \gls{soc} architecture described in \Cref{sec:baseline-soc}.
We extend the \gls{dma} engine and Snitch's \gls{lsu} to inject the collective opcode in the \texttt{AWUSER} field of their outgoing AXI requests.

\subsubsection{\acrfull{dca}}
\label{sec:dca}

We further extend the Snitch cluster to support \acrfull{dca}, a novel paradigm that enables direct access to the cluster's compute resources.
Analogous to how \gls{dma} engines can directly access memory while the cores perform other tasks, \gls{dca} grants the interconnect fabric direct access to the cores' compute resources, e.g. to carry out in-network computations, while the cores perform other work or enter a low-power state to save energy.

To support \gls{dca}, we equip the Snitch cluster with three 512-bit ports: two for input operands and one for the result.
Additional control signals on the interface specify the operation type.
Inside the cluster, each 512-bit operand is divided into eight 64-bit slices, which are distributed to the respective cores' \glspl{fpu} for parallel processing.
Within the Snitch core complex, \gls{dca} requests are arbitrated with the core's own \gls{fpu} requests.
A tag is used to differentiate \gls{dca} and core requests, propagated through the \gls{fpu}'s pipeline stages and used to route the result to the correct destination.

We connect the \gls{dca} interface of the Snitch cluster to the router's offload interface, reusing the existing datapath to perform wide in-network reduction operations with a negligible overhead, as shown in \Cref{sec:at-results}.
With the SIMD capabilities of Snitch's \glspl{fpu}, the system can perform up to 8$\times$ double-precision or 64$\times$ 8-bit precision floating-point reductions per cycle.

The operands sent by the router can be buffered within the pipeline registers (\lstinline{cut}) between the router and the arithmetic units and all paths are further provided with independent valid-ready interfaces, which allow the arithmetic unit to exert backpressure on an operand to wait for the other.

\subsubsection{System Address Map}
\label{sec:address-map}

As described in \Cref{sec:encoding}, the adopted multi-address encoding scheme \cite{colagrande2025} trades flexibility for scalability, imposing specific constraints on the system design.
In particular, the collective-targetable region of the \gls{noc} must form a submesh defined by parameters $(X, Y, W, H)$, where $W$ and $H$ specify the width and height of the submesh, and $(X, Y)$ denotes the coordinates of its bottom-left tile.
To maintain compatibility with the encoding scheme, the following conditions must hold:
1) both $W$ and $H$ must be powers of two, and
2) $X$ and $Y$ must be aligned to integer multiples of $W$ and $H$, respectively.
In most cases, these conditions can be satisfied by ``padding'' the mesh, i.e. by artificially aligning the collective-targetable region, as shown in \Cref{fig:soc}.

To further simplify the translation logic from address mask to X and Y masks, we assume that the address space of all nodes in the collective-targetable region is
1) of equal size,
2) aligned to the same power of two, and
3) mapped consecutively following the Y-major ordering of the node coordinates.
Under these assumptions, the translation reduces to an efficient bit-select operation on the address mask.

\subsection{Generalizability}
\label{sec:generalizability}

We demonstrate our implementation on a Snitch cluster and FlooNoC-based system as it represents an open-source instance of a common architectural template.
In fact, the proposed approach does not rely on any Snitch- or FlooNoC-specific mechanims, but instead builds on three general architectural features shared by a broad class of accelerators:
\begin{enumerate}[
  itemsep=0pt,
  topsep=2pt,
  parsep=0pt,
  partopsep=0pt
]
    \item Structured 2D mesh topology: the XY routing and coordinate-based multi-address encoding (\Cref{sec:encoding}) require a regular mesh topology (\Cref{sec:address-map}).
    \item Arithmetic units: the \gls{dca} paradigm requires each compute tile to expose arithmetic units that can be borrowed to perform the in-network reductions%
    \footnote{While in our implementation the cluster's 8$\times$ 64-bit \glspl{fpu} match the 512-bit data width of the wide network and the \gls{dca} interface, the same mechanism could be extended to narrower or wider (through time-sharing) interfaces than the compute datapath.}.
    \item Programmable communication: the proposed mechanism assumes that communication and data movement can be orchestrated programmatically, independent of the specific control interface or data movement engine, e.g. \gls{dma}, tensor streaming engine or \gls{lsu}.
\end{enumerate}

These features are characteristic of many recent academic and industrial programmable \gls{ml} accelerators.
To name a few, Cerebras WSE-3~\cite{cerebras_wse3}, Tenstorrent's Blackhole~\cite{tenstorrent_blackhole}, AMD's XDNA~\cite{amd_xdna}, SambaNova's SN40L~\cite{sambanova2024} and Meta's MTIA~\cite{metamtia2023} among industrial accelerators, and Venus~\cite{venus2023}, Adyna~\cite{adyna2025}, FlatAttention~\cite{zhang2025}, MAGIA~\cite{fractalsync2025}, and Azul~\cite{azul2024} among academic accelerators, all share a regular 2D tiled topology, arithmetic resources embedded in each tile and programmable engines for orchestrating communication and data movement, either at the tile or global level.

\section{Results}
\label{sec:results}

\subsection{Area and Timing Analysis}
\label{sec:at-results}

We implement the \gls{ni} and router modules, as well as a full cluster tile, in \textsc{TSMC} \SI{7}{\nano\meter} technology using \textsc{Fusion Compiler 2024.09}, with a \SI{1}{\giga\hertz} frequency target under worst-case conditions (SS, $-40^\circ$C, \SI{0.675}{\volt}), observing no timing degradations.

Since the extensions to the \gls{ni} are identical regardless of which collective operations are supported, we compare the baseline \gls{ni} with the version featuring full collective support, which adds only a 3.5\% area overhead.

\begin{figure}[t]
    \centering
    \begin{subfigure}{0.52\columnwidth}
        \centering
        \includegraphics[width=\textwidth]{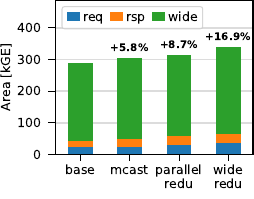}
        \vspace{-1.6em}
        \caption{}
        \label{fig:router-area}
    \end{subfigure}
    \hfill
    \begin{subfigure}{0.46\columnwidth}
        \centering
        \includegraphics[width=\textwidth]{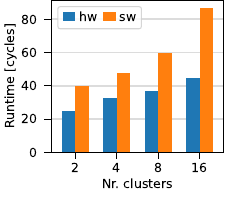}
        \vspace{-1.6em}
        \caption{}
        \label{fig:barrier}
    \end{subfigure}
    \vspace{-0.6em}
    \caption{(a) Area breakdown of the router for different hardware configurations. Percentages indicate the area overhead with respect to the baseline. (b) Runtime of the software and hardware barriers.}
    \label{fig:area-and-barrier}
    \vspace{-1em}
\end{figure}

To assess the impact of individual collective operations on the router area, we compare the baseline router with multiple configurations of the collective-capable router, progressively adding support for multicast, parallel reductions and wide reductions.
\Cref{fig:router-area} reports the area of all configurations.

Adding \texttt{multicast} support introduces flit-forking logic in both the narrow and wide routers, resulting in only a \ResultMulticastForkingOverhead\% area overhead compared to the baseline.
As described in \Cref{sec:impl-parallel-reduction}, minimal support for parallel reduction is also required in the response router to merge responses from multiple subordinates.
This resource accounts for \ResultMulticastCollectBOverhead\% of the response router area, leading to a total router area overhead of just \ResultMulticastRouterOverhead\% for full multicast support.

Enabling \texttt{parallel} reduction for synchronization mechanisms increases the router area by only \ResultParallelRedRouterOverhead\%.
The main addition lies in the narrow request router, where the reduction arbiters combining the incoming flit data introduce \SI{\ResultRedArbiterGe}{\kilo\ge} per output port.
Because of the coupling between multicast and reduction, enabling reduction also requires forking logic in the response router to multicast responses.

Adding wide reduction support introduces an additional \SI{\ResultWideSimpleControllerGe}{\kilo\ge} overhead, primarily due to the increased data width.
\ResultWideSimpleCrtlComb\% of this logic is combinational and \ResultWideSimpleCrtlSeq\% sequential, the former dominated by multiplexers for input arbitration and the latter by the flit header buffer.
Overall, full support for all collective communication operations described in \Cref{sec:implementation} results in a modest \ResultAreaOverheadTot\% area increase over the baseline router.

\begin{figure}[t]
    \centering
    \includegraphics[width=0.6\columnwidth]{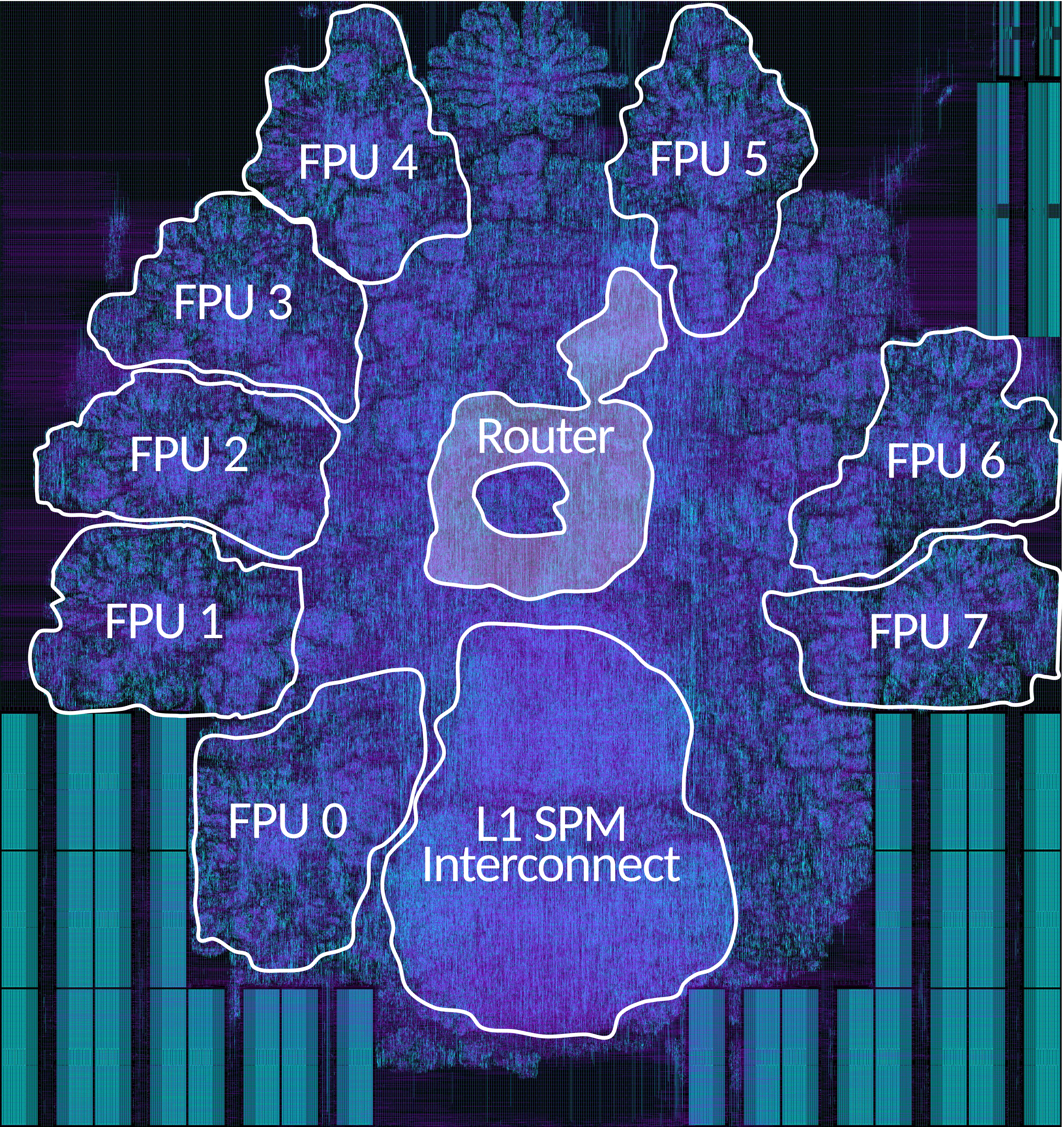}
    \caption{Placed-and-routed implementation of the cluster tile, with the \glspl{fpu}, the router and the L1 \gls{spm} interconnect highlighted. The remaining area is occupied by the Snitch cores, L1 \gls{spm}, I\$ subsystem and cluster \gls{dma}, which are not highlighted for clarity.}
    \label{fig:floorplan}
\end{figure}

To better quantify the impact of our approach at the system level, we reproduce the place-and-route flow of the cluster tile from ~\cite{fischer2025}.
\Cref{fig:floorplan} highlights the area occupied by the FlooNoC router and the cluster's \glspl{fpu}.
As we can see, the \glspl{fpu} occupy a significant portion of the tile area, well beyond the area of the router, highlighting the significance of the \gls{dca} paradigm in enabling high-throughput in-network reductions at low hardware cost.
Compared to the \SI{5.6}{\mega\ge} cluster tile area, the overhead introduced by our extensions is negligible, falling below \textless\,1\%.

\subsection{Performance Evaluation of Collective Primitives}
\label{sec:collectives-performance}

We conduct the performance evaluation through cycle-accurate RTL simulations of the system described in \Cref{sec:baseline-soc,sec:impl-system}, using \textsc{QuestaSim 2023.4}.

All benchmark codes are implemented in bare-metal C++, compiled using a custom C++ runtime and compiler toolchain for Snitch based on LLVM 15 with highest optimization level (-O3), and further optimized by hand to reduce or hide the overhead of non-communication related instructions.
For transparency and accountability, all codes are open source%
\footnote{\ifdefined\blindreview
    \url{https://hidden-for-double-blind-review.com}
\else
    \url{https://github.com/pulp-platform/picobello/blob/cacdc3ad4a3a638e68d03090781dbbc2450c73ed/MLSYS.md}
\fi}.

\subsubsection{Narrow Reduction}
\label{sec:narrow-reduction-results}

We assess the benefit of hardware-accelerated narrow reduction operations using a simple yet fundamental parallel programming primitive: barrier synchronization.
Prior studies have shown that parallel performance is not only limited by sequential code, as Amdahl's law suggests \cite{amdahl1967}, but also by synchronization overheads \cite{yavits2014, eyerman2010}.
Accelerating barrier synchronization can therefore improve parallel performance, as we further demonstrate in \Cref{sec:wide-multicast-results,sec:wide-reduction-results}.

Typical software implementations of the barrier primitive rely on all participants atomically incrementing a centralized counter \cite{culler1998}, which we allocate in cluster 0's L1 \gls{spm}.
For an efficient and scalable implementation, we use the RISC-V \lstinline{amoadd} atomic instruction: every atomic operation arriving at the destination memory completes with a latency of 3 cycles, that is one cycle for each of the 1) read, 2) modify and 3) write operations implied by the atomic operation.
Cluster interrupts are used to notify all participants of barrier completion, avoiding busy-waiting, and in-network multicast support is leveraged to distribute interrupts simultaneously.
Barrier runtime is measured from the arrival of the first core to the departure of the last.

We compare this software baseline against a barrier implemented using in-network reductions.
In this version, all participants perform an \lstinline{LsbAnd} reduction on the central counter, followed by a RISC-V \lstinline{fence} instruction, which stalls the core until all memory operations complete;
reduction operations only complete once all cores have sent their contribution, i.e. once they have arrived on the barrier.

\Cref{fig:barrier} presents the results of this comparison, for varying numbers of clusters participating in the barrier.
Although both implementations scale linearly with the number of clusters, the hardware-assisted barrier scales significantly better, as packets are reduced in the network along their path to the destination, avoiding wasteful read-modify-write cycles.
Through linear regression, we find slopes of \ResultSlopeSoftwareBarrier\ and \ResultSlopeHardwareBarrier\ cycles per additional cluster for the software and hardware barriers, respectively, which match well the expected values of 3 and 1 cycles per cluster.

\subsubsection{Wide Multicast}
\label{sec:wide-multicast-results}

While the narrow network primarily transports short, latency-critical synchronization messages, the wide network transfers large data bursts initiated by the \gls{dma} engines.
These transfers typically overlap with the computation in a double-buffered fashion \cite{potocnik2024, zhang2025}.
Optimizing transfer time can substantially improve performance in memory-bound workloads.
When spatial data reuse is present, multicast can be employed to this end, as demonstrated in \Cref{sec:gemm-results}.

To evaluate the benefit of hardware-accelerated multicast transfers, we compare against the runtime of two optimized software baselines.
Both baselines employ the \gls{dma} engines to orchestrate data transfers and adopt the hardware-accelerated barrier from \Cref{sec:narrow-reduction-results}, leveraging the hardware reduction extension in the narrow router, for inter-cluster synchronization.
To simplify the analysis, we begin with a 1D multicast along a single row, moving data from memory tile 0 to the L1 \gls{spm} of every cluster in row 0.

\begin{figure}[t]
    \centering
    \includegraphics[width=\columnwidth]{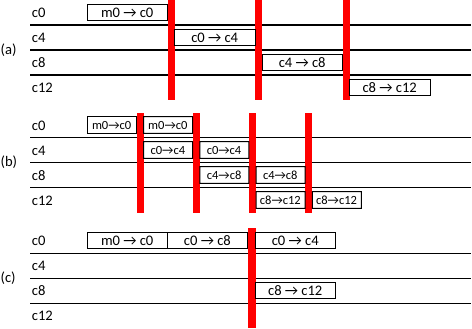}

    \begingroup
        \phantomsubcaption
        \label{fig:naive-mcast}
        \phantomsubcaption
        \label{fig:seq-mcast}
        \phantomsubcaption
        \label{fig:tree-mcast}
    \endgroup

    \caption{Three software multicast implementations: (a) naive sequential, (b) pipelined sequential, (c) tree-based. Each block represents a \gls{dma} transfer: the containing row represents the initiator and the label indicates source and destination (source $\rightarrow$ destination). Red lines represent barriers.}
    \label{fig:1d-mcast}
    \vspace{-1em}
\end{figure}

In the first baseline, each cluster fetches data from its left neighbour once that neighbour has completed its own transfer.
As shown in \Cref{fig:naive-mcast}, each transfer starts only after the full transfer in the previous iteration is complete, requiring synchronization among clusters at each iteration.
Given $c$ clusters in a row, the runtime can be modeled by:
\begin{equation}
    \label{eq:naive-mcast}
    T_{naive} = \sum_{i=1}^{c} (\alpha_i + \beta n + \delta) - \delta
\end{equation}
where $\alpha_i$ is the round-trip latency of the \gls{dma} transfer in each iteration $i$, $\beta$ the inverse bandwidth (cycles per beat), $n$ the transfer size (in beats) and $\delta$ the synchronization time.
\Cref{fig:seq-mcast} illustrates an optimized implementation (\lstinline{seq}) dividing the transfer into $k$ batches that are pipelined across clusters.
The runtime then becomes:
\begin{equation}
    \label{eq:seq-mcast}
    T_{seq} = \sum_{i=1}^{k + c - 1} (\alpha_i + \frac{n}{k} \beta + \delta) - \delta
\end{equation}
As this equation suggests, an optimal batch size exists that minimizes overall transmission time.
Intuitively, this is due to the additional round-trip latency ($\alpha$) and synchronization ($\delta$) overheads offsetting the benefit from the decreased batch size $\frac{n}{k}$.
Our second baseline (\lstinline{tree}) implements a binary-tree multicast, as depicted in \Cref{fig:tree-mcast}, modeled as:
\begin{equation}
    \label{eq:tree-mcast}
    T_{tree} = \sum_{i=0}^{\log_2 c} (\alpha_i + n \beta + \delta) - 2 \delta
\end{equation}

Both implementations have been extensively studied and detailed comparisons can be found in the networking literature \cite{barnett1996}.
We focus on the comparison with hardware-accelerated multicast\footnote{We do not consider a pipelined tree-based implementation, as simultaneous transfers of different batches cross the same physical links and result in contention, eliminating any benefit.}, with runtime:
\begin{equation}
    \label{eq:hw-mcast}
    T_{hw} = \alpha + (n + c - 1) \beta
\end{equation}

\begin{figure}[t]
    \begin{subfigure}{\columnwidth}
        \centering
        \includegraphics[width=\textwidth]{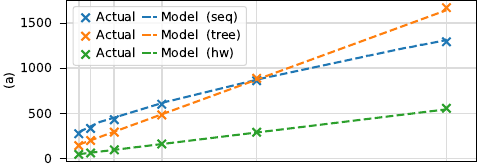}
        \phantomcaption
        \label{fig:mcast-results}
    \end{subfigure}
    \vspace{-1em}

    \begin{subfigure}{\columnwidth}
        \centering
        \includegraphics[width=\textwidth]{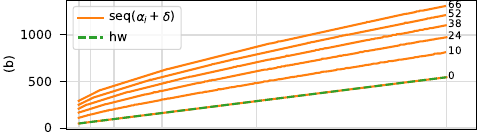}
        \phantomcaption
        \label{fig:seq-convergence}
    \end{subfigure}
    \vspace{-1em}

    \begin{subfigure}{\columnwidth}
        \centering
        \includegraphics[width=\textwidth]{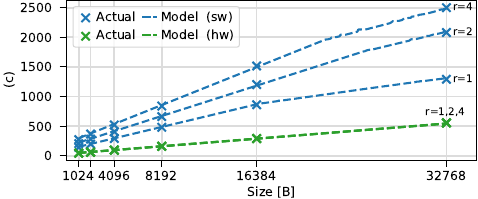}
        \phantomcaption
        \label{fig:2d-mcast-results}
    \end{subfigure}
    \vspace{-1.4em}

    \caption{Runtime (in cycles) of: (a) a 1D multicast transfer; (b) the \lstinline{seq} implementation for various settings of $\alpha_i + \delta, \forall\ i > 0$, labeled next to each curve; (c) a 2D multicast transfer.}
    \label{fig:all-mcast-results}
    \vspace{-0.6em}
\end{figure}

\Cref{fig:mcast-results} presents the results of this comparison.
All runtimes are measured from the start of the first \gls{dma} transfer, to the end of the last, and the optimal batch size is assumed for the \lstinline{seq} implementation.
As can be seen, all models accurately reflect the measured runtimes.
The hardware multicast implementation consistently outperforms both software baselines, with speedups between \ResultMinHwMcastSpeedup$\times$ and \ResultMaxHwMcastSpeedup$\times$ over the best software baseline, $T_{sw}=\min(T_{seq}, T_{tree})$, for all transfer sizes between \SI{\ResultMinMcastSizeKibiBytes}{} and \SI{\ResultMaxMcastSizeKibiBytes}{\kibi\byte}.

Furthermore, we note that the \lstinline{hw} implementation can be viewed as a degenerate case of the \lstinline{seq} implementation, in which the transfers are overlapped with a granularity of a single beat ($k=n$), and without incurring any overhead from splitting the transfer in batches ($\alpha_i=0, \forall\,i > 0$ and $\delta=0$).
\Cref{fig:seq-convergence} illustrates this behaviour, showing how $T_{seq}$ converges to $T_{hw}$ as $\alpha_i + \delta$ approaches 0.
This highlights the importance of minimizing the barrier synchronization overhead ($\delta$), as shown in \Cref{sec:narrow-reduction-results}.

\Cref{fig:1d-mcast} and \Crefrange{eq:naive-mcast}{eq:hw-mcast} can be easily generalized to a multicast transfer involving multiple rows, as presented in \Cref{sec:appendix-2d-mcast}.
In short, a 2D multicast transfer can be implemented in two steps: 1) a first 1D multicast distributes the data across one row, and 2) $c$ independent 1D transfers then distribute the data to each column, in parallel.
\Cref{fig:2d-mcast-results} compares the runtime of the \lstinline{hw} and best software implementation for varying numbers of rows $r$.
While the runtime of the software implementations significantly increases with the number of rows, the hardware implementation remains nearly constant, \textit{enabling scalable and efficient multicast operations on large networks}.
Overall, we measure a geomean speedup of \ResultGeomeanHwMcastTwoDSpeedup$\times$ on broadcasts between \SI{\ResultMinMcastSizeKibiBytes}{} and \SI{\ResultMaxMcastSizeKibiBytes}{\kibi\byte} in the 4$\times$4 mesh.

\subsubsection{Wide Reduction}
\label{sec:wide-reduction-results}

Likewise, reduction operations can also benefit from in-network acceleration.
As discussed in \Cref{sec:gemm-results}, such operations are common in \gls{ml} workloads, and accelerating them can greatly improve application performance.

We compare hardware-accelerated reductions against two optimized software baselines, starting with a 1D reduction.
As in \Cref{sec:wide-multicast-results}, both baselines use the \glspl{dma} and the hardware-accelerated barrier for synchronization.
In this operation, each cluster contributes a stream of data from its L1 \gls{spm} to be elementwise reduced with all other clusters' data into a centralized destination, e.g. cluster 0.

\begin{figure}[t]
    \centering
    \includegraphics[width=\columnwidth]{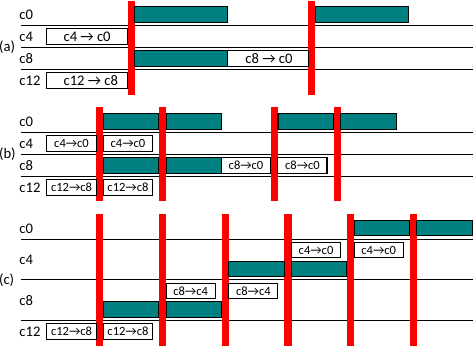}

    \begingroup
        \phantomsubcaption
        \label{fig:naive-tree-reduction}
        \phantomsubcaption
        \label{fig:tree-reduction}
        \phantomsubcaption
        \label{fig:seq-reduction}
    \endgroup

    \caption{Three software reduction implementations: (a) naive tree, (b) double-buffered tree, (c) pipelined sequential. Each block represents a \gls{dma} transfer: the containing row represents the initiator and the label indicates source and destination (source $\rightarrow$ destination). Red lines represent barriers. Colored blocks represent the reduction computations.}
    \label{fig:1d-reduction}
    \vspace{-0.5em}
\end{figure}

The software implementations of the reduction operation, represented in \Cref{fig:1d-reduction}, resemble the multicast baselines presented in \Cref{sec:wide-multicast-results}, with two substantial differences: 1) the flow of the data is mirrored (one-to-many for multicast vs. many-to-one for reduction), and 2) reduction involves not only data movement but also computation.
We further optimize the naive tree implementation shown in \Cref{fig:naive-tree-reduction} by tiling the computation to overlap data movement and computation, as illustrated in \Cref{fig:tree-reduction}.
The runtime of the optimized tree (\lstinline{tree}) and sequential (\lstinline{seq}) implementations can be modeled as:
\begin{equation}
    \label{eq:seq-reduction}
    \begin{split}
        T_{seq} & = t_m + 2(c - 2)\max(t_m, t_c) + k t_c + \\
                & + (2 (c - 2) + k) \delta
    \end{split}
\end{equation}
\begin{equation}
\label{eq:tree-reduction}
    \begin{split}
        T_{tree} & = \left\{ t_m + \delta + (k - 1)\left[\max(t_m, t_c) + \delta\right] + \right. \\
                 & \left. + t_c \right\} * \log_2 c
    \end{split}
\end{equation}
with $t_{m} = \alpha_{m} + \frac{n}{k} \beta_{m}$ and $t_{c} = \alpha_{c} + \frac{n}{k} \beta_{c}$, given $\alpha_m$ and $\beta_m$ the round-trip latency and inverse bandwidth of the \gls{dma} transfers, and $\alpha_c$ and $\beta_c$ respectively a constant instruction overhead and the inverse throughput (cycles per beat) of the computation.
As seen for multicast, tiling the transfers introduces additional synchronization and round-trip latency overheads, leading to the existence of an optimal tile size which minimizes the transmission time.

\begin{figure}[t]
    \begin{subfigure}{\columnwidth}
        \centering
        \includegraphics[width=\textwidth]{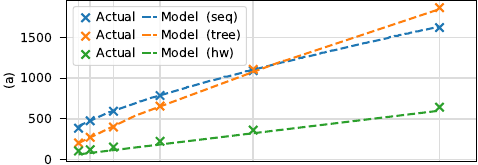}
        \phantomcaption
        \label{fig:reduction-results}
    \end{subfigure}
    \vspace{-1em}

    \begin{subfigure}{\columnwidth}
        \centering
        \includegraphics[width=\textwidth]{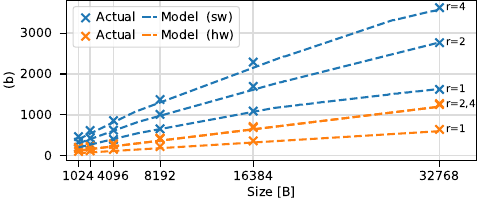}
        \phantomcaption
        \label{fig:2d-reduction-results}
    \end{subfigure}
    \vspace{-1.4em}

    \caption{Runtime (in cycles) of: (a) a reduction transfer of varying size to one row of clusters; (b) a 2D reduction transfer.}
    \label{fig:all-reduction-results}
    \vspace{-1em}
\end{figure}

\begin{figure*}[t]
    \centering
    \begin{subfigure}{0.475\textwidth}
        \centering
        \includegraphics[width=\textwidth]{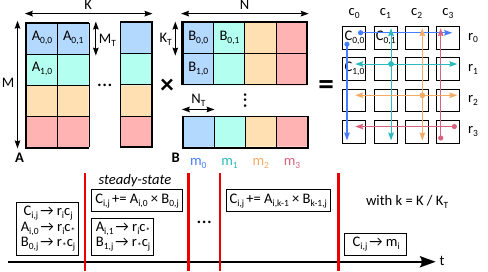}
        \caption{}
        \label{fig:gemm-mcast-dataflow}
    \end{subfigure}
    \hfill
    \begin{subfigure}{0.515\textwidth}
        \centering
        \includegraphics[width=\textwidth]{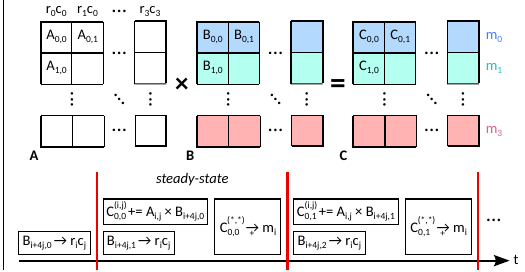}
        \caption{}
        \label{fig:gemm-reduction-dataflow}
    \end{subfigure}
    \vspace{-0.6em}
    \caption{\gls{gemm} dataflows mapped onto a 4$\times$4 tile-based architecture: (a) SUMMA \gls{gemm} \cite{vandegeijn1995}; (b) FusedConcatLinear \gls{gemm} \cite{potocnik2024}. Colored background indicates L2 storage location (blue: \lstinline{m0}, teal: \lstinline{m1}, yellow: \lstinline{m2}, red: \lstinline{m3}). Colored arrows illustrate data movement: the tail marks the \textit{initiator}, the color the \textit{source} L2 tile, and the traversed clusters are the \textit{destinations}. In the timing diagrams: red lines indicate barriers; $i$ and $j$ denote row and column indices, respectively. The ($Y \rightarrow x$) notation denotes a data transfer of matrix tile $Y$ to destination $x$: stars imply collective operations (multicast if on the right-hand side, reduction if on the left-hand side), with an operator on the arrow specifying the reduction operation.}
    \label{fig:gemm-dataflow}
\end{figure*}

\Cref{fig:reduction-results} presents the comparison to hardware-accelerated reduction.
Again, our models accurately reflect the measured runtimes.
The hardware implementation consistently outperforms both baselines, achieving speedups between \ResultMinHwReductionSpeedup$\times$ and \ResultMaxHwReductionSpeedup$\times$ over the best software baseline, $T_{sw}=\min(T_{seq}, T_{tree})$, for transfer sizes between \SI{\ResultMinReductionSizeKibiBytes}{} and \SI{\ResultMaxReductionSizeKibiBytes}{\kibi\byte}.

\Cref{fig:1d-reduction} and \Crefrange{eq:seq-reduction}{eq:tree-reduction} can be easily generalized to a reduction involving multiple rows, as presented in \Cref{sec:appendix-2d-reduction}.
In short, a 2D reduction can be implemented in two steps: 1) $c$ independent 1D reduction operations combine the data across each row, in parallel, and 2) a final column-wise 1D reduction combines these partial results.

\Cref{fig:2d-reduction-results} shows a comparison between the runtime of the \lstinline{hw} implementation and the best software implementation for a 2D reduction involving varying number of rows $r$.
Differently from the multicast case, the runtime of the hardware implementation increases significantly when going from a 1D to a 2D reduction.
As only 2 inputs can be combined at a time, reducing more than 2 inputs requires multiple cycles.
This is the case at the routers of the first column (excluding the northern-most router), which receive data from three inputs (specifically the east, north and local inputs).
This results in a throughput of only one fully-reduced beat every 2 cycles, explaining the \ResultOneDimToTwoDimSlowdown$\times$ slowdown of the \SI{\ResultOneDimToTwoDimSlowdownKibiBytes}{\kibi\byte} transfer.
Nonetheless, while the software implementations scale poorly with the number of rows, the runtime of the hardware implementation remains near constant beyond two rows, thus \textit{enabling scalable and efficient reduction operations on large networks}.
Overall, we measure a geomean speedup of \ResultGeomeanHwReductionTwoDSpeedup$\times$ on reductions between \SI{\ResultMinReductionSizeKibiBytes}{} and \SI{\ResultMaxReductionSizeKibiBytes}{\kibi\byte} in the 4$\times$4 mesh.

\subsection{Evaluation of GEMM Kernels}
\label{sec:gemm-results}

In this section, we demonstrate and explain how multicast and reduction acceleration can translate to actual kernel speedups.
We illustrate these concepts through the use of two examples, targeting \gls{gemm}, a key kernel for \gls{ml} workloads.
For this evaluation, we develop analytical models of the \gls{gemm} runtime, as a composition of empirically-validated models: those developed in \Cref{sec:collectives-performance} for $T_{comm}$, and a model developed in previous work for $T_{comp}$.

\subsubsection{SUMMA GEMM}

Consider a \gls{gemm} operation $C = \alpha A \times B + \beta C$, with matrices $A$, $B$ and $C$ respectively of size $M \times K$, $K \times N$ and $M \times N$.
We map the \gls{gemm} computation on the system described in \Cref{sec:baseline-soc,sec:impl-system}, employing the SUMMA dataflow \cite{vandegeijn1995}, with double-buffering to overlap computation and communication, as illustrated in \Cref{fig:gemm-mcast-dataflow}.
In every iteration, each cluster computes a subproblem of size $M_t \times N_t \times K_t$.
To benefit from data reuse at the L2 memory level, we assume that the memory tiles are dimensioned to jointly fit a maximum problem size $M \times N \times K$ that satisfies either one of the following conditions: $M \gg r M_t$, $N \gg c N_t$ and $K \gg K_t$.
For the sake of example, we assume $K \gg K_t$, $M = r M_t$ and $N = c N_t$, but analogous arguments can be made for other configurations.

Under the $K \gg K_t$ assumption, we can ignore boundary iterations and focus on a steady-state iteration's runtime:
\begin{equation}
    \label{eq:gemm-runtime}
    T = \max(T_{comp}, T_{comm})
\end{equation}
\begin{equation}
    \label{eq:gemm-compute-time}
    T_{comp} = \frac{2 M_t N_t K_t}{\variable{Util} \cdot \variable{PeakPerf}}
\end{equation}
\begin{equation}
    \label{eq:gemm-communication-time}
    T_{comm} = T_{mcast_A} + T_{mcast_B}
\end{equation}
where $T_{mcast_A}$ and $T_{mcast_B}$ are the times required to transfer a submatrix $A_{i,*}$ to all clusters in row $i$ and a submatrix $B_{*,j}$ to all clusters in column $j$, respectively.

We use \Crefrange{eq:seq-mcast}{eq:hw-mcast} to model these transfers, selecting the best software implementation on a case-by-case basis, and evaluate the effect of supporting in-network multicast operations on \gls{gemm} performance.
\Cref{fig:gemm-mcast-results} shows the results of this evaluation for different mesh sizes, assuming the maximum square problem size fitting in a cluster L1 \gls{spm} of \SI{16}{\kibi\byte}, and a \ResultMedianGemmUtilizationPercentage\% utilization%
\footnote{Median utilization across various \gls{gemm} sizes reported by \cite{colagrande2025b}, measured through cycle-accurate RTL simulation. As such, it accounts for real-world effects, such as memory contention between the DMA and the compute cores (the bottleneck they address), and detailed instruction-level scheduling.}.
By accelerating multicast in hardware, the operation stays compute-bound up to a notable 256$\times$256 mesh size, while the software implementation becomes memory-bound already on a 16$\times$16 mesh, resulting in speedups between \ResultMinGemmHwMcastSpeedup$\times$ and \ResultMaxGemmHwMcastSpeedup$\times$.

\subsubsection{FusedConcatLinear GEMM}

To evaluate reductions, we focus on the use-case illustrated by \cite{potocnik2024}, though other works have also demonstrated the benefit of fast reductions on \gls{ml} workloads \cite{zhang2025}.
Consider a \gls{mha} layer \cite{vaswani2017}, where each cluster is assigned the computation of a distinct attention head.
\cite{potocnik2024} demonstrate that by fusing the final concatenation and linear layers with the attention computations, costly external memory accesses can be avoided.
This scheme boils down to a \gls{gemm} distributed across clusters along the $K$ dimension, as illustrated in \Cref{fig:gemm-reduction-dataflow}.
As a result, a final reduction operation is required to aggregate the partial results of $C$ from all clusters.

\begin{figure}[t]
    \begin{subfigure}{\columnwidth}
        \centering
        \includegraphics[width=\textwidth]{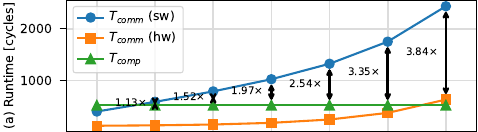}
        \phantomcaption
        \label{fig:gemm-mcast-results}
    \end{subfigure}
    \vspace{-1em}

    \begin{subfigure}{\columnwidth}
        \centering
        \includegraphics[width=\textwidth]{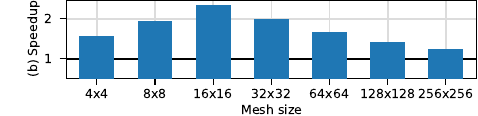}
        \phantomcaption
        \label{fig:gemm-reduction-results}
    \end{subfigure}
    \vspace{-1.8em}

    \caption{(a) Runtime of the communication and computation phases of the SUMMA \gls{gemm} kernel. (b) Hardware vs. software reduction speedup for the FusedConcatLinear \gls{gemm} kernel. The X-axis uses a logarithmic (base 2) scale.}
    \label{fig:all-gemm-energy}
    \vspace{-2em}
\end{figure}

\begin{table*}[t]
    \caption{Energy cost of primitive operations and data movement/compute counts per \gls{gemm} implementation on a $16{\times}16$ mesh. Transfer counts are in kilo-bytes [\si{\kilo\byte}] and compute counts in kilo-operations [\si{\kilo OP}].}
    \label{tbl:gemm-energy}
    \vspace{3pt}
    {\setlength{\tabcolsep}{-4pt}
    \begin{tabularx}{\textwidth}{@{} l *{7}{>{\centering\arraybackslash}X} @{}}
        \toprule
        Primitives & \gls{dma} Load & \gls{dma} Store & Hop & \gls{spm}~Write & \gls{gemm} & SW Reduce & \mbox{DCA Reduce} \\
        \midrule
        Energy           & \SI{\ResultEnDmaLd}{\pico\joule/\byte} & \SI{\ResultEnDmaSt}{\pico\joule/\byte} & \SI{\ResultEnLinkToLink}{\pico\joule/\byte} & \SI{\ResultEnWriteTcdm}{\pico\joule/\byte} & \SI{\ResultEnGemm}{\pico\joule/OP} & \SI{\ResultEnSwRed}{\pico\joule/OP} & \SI{\ResultEnHwRed}{\pico\joule/OP}\rlap{\tblmark{3}} \\
        \midrule
        SUMMA SW & \ResultSummaSwDmaLd & \ResultSummaSwDmaSt & \ResultSummaSwLinkToLink & \ResultSummaSwTcdmWrite & \ResultSummaSwGemm & \ResultSummaSwSwRed & \ResultSummaSwHwRed \\
        SUMMA HW & \ResultSummaHwDmaLd & \ResultSummaHwDmaSt\tblmark{1} & \ResultSummaHwLinkToLink & \ResultSummaHwTcdmWrite & \ResultSummaHwGemm & \ResultSummaHwSwRed & \ResultSummaHwHwRed \\
        FCL SW   & \ResultFclSwDmaLd   & \ResultFclSwDmaSt   & \ResultFclSwLinkToLink   & \ResultFclSwTcdmWrite   & \ResultFclSwGemm   & \ResultFclSwSwRed   & \ResultFclSwHwRed   \\
        FCL HW   & \ResultFclHwDmaLd   & \ResultFclHwDmaSt\tblmark{2}   & \ResultFclHwLinkToLink   & \ResultFclHwTcdmWrite\tblmark{2}   & \ResultFclHwGemm   & \ResultFclHwSwRed   & \ResultFclHwHwRed   \\
        \bottomrule
    \end{tabularx}}
    \vspace{-5pt}
\end{table*}

\Cref{fig:gemm-reduction-results} shows the results of this evaluation.
While the trend differs from the multicast case, these results similarly highlight the benefit of accelerating reductions in hardware, demonstrating up to \ResultMaxGemmHwReductionSpeedup$\times$ speedups on a key computational kernel for \gls{ml} workloads, in the evaluated scenario%
\footnote{In general, \gls{dca} requests and regular \gls{fpu} instructions compete for the same arithmetic units (\Cref{sec:dca}), potentially creating contention that complicates performance reasoning when the two phases overlap in time. In the FusedConcatLinear \gls{gemm}, however, the reduction phase strictly follows the computation phase, so no such contention arises.}.

\subsubsection{GEMM Energy}

To evaluate the impact of our extensions on energy, we estimate the energy consumption of both GEMM workloads, with and without hardware-accelerated collectives, across different mesh sizes.
We perform gate-level simulations of the full-system mesh, after replacing cluster tile 0 with its post-layout netlist.
We then use \textsc{PrimeTime 2022.03} to estimate the energy of cluster tile 0, from the switching activity extracted during simulation, in the typical corner (TT, $25^\circ$C, \SI{0.75}{\volt}) with a \SI{1}{\giga\hertz} clock frequency.

We measure the energy of the primitive operations reported in \Cref{tbl:gemm-energy}.
We then break down each GEMM workload into its constituent operations, and count the number of occurrences of each primitive across all cluster tiles (reported in \Cref{tbl:gemm-energy} for a $16{\times}16$ mesh), to obtain an estimate of the total energy%
\footnote{
    We do not account for the energy spent in accesses to the L2 memory, as this is the same across both implementations.
}.
The software-based implementation always assumes the fastest software collective as in \Cref{sec:gemm-results}.

\begin{figure}[t]
    \begin{subfigure}{\columnwidth}
        \centering
        \includegraphics[width=\textwidth]{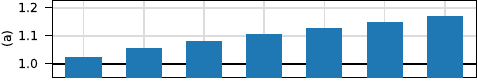}
        \phantomcaption
        \label{fig:gemm-summa-energy}
    \end{subfigure}
    \vspace{-1.5em}

    \begin{subfigure}{\columnwidth}
        \centering
        \includegraphics[width=\textwidth]{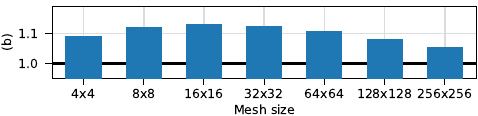}
        \phantomcaption
        \label{fig:gemm-fcl-energy}
    \end{subfigure}
    \vspace{-2.4em}

    \caption{Energy saving of the hardware-accelerated over the software-based \gls{gemm} implementations for the SUMMA (a) and FusedConcatLinear (b) dataflows across different mesh sizes.}
    \label{fig:all-reduction-results}
    \vspace{-1em}
\end{figure}

\Cref{fig:gemm-summa-energy} reports the energy savings on the SUMMA \gls{gemm} kernel across different mesh sizes.
As quantified in \Cref{tbl:gemm-energy}\tblmark{1}, hardware multicast reduces the number of DMA operations involved in the multicast transfers of $A_{i,*}$ and $B_{*,j}$ submatrices.
While the total energy consumption remains dominated by computation, the communication savings grow with the mesh size, reaching overall energy efficiency improvements of up to \ResultSummaMaxEnergySaving$\times$ for a $256 \times 256$ mesh, and demonstrating the increasing benefit of multicast support with scale.

\Cref{fig:gemm-fcl-energy} shows the energy savings enabled by reduction support for the FusedConcatLinear \gls{gemm}.
The in-network reduction and \gls{dca} paradigm yield energy improvements of up to \ResultFclMaxEnergySaving$\times$.
These gains stem from two main factors.
First, the baseline software implementation requires explicit inter-cluster data movement to accumulate partial results, increasing communication energy\tblmark{2}.
Second, without \gls{dca}, the baseline must keep all eight Snitch cores active to initiate \gls{fpu} operations.
On the other hand, with our approach the \gls{fpu} operations are initiated directly by the \gls{dca} requests, allowing the cores to remain in a low-power state\tblmark{3}.

\subsubsection{General Observations}

While our evaluation is limited to two \gls{gemm} kernels, this selection serves to illustrate the general, kernel-independent conditions for hardware collective acceleration to translate to tangible kernel speedups:
\begin{enumerate}[
  itemsep=0pt,
  topsep=0pt,
  parsep=0pt,
  partopsep=0pt
]
\item communication must be on the critical path of the kernel's runtime and
\item the communication pattern must map to multicast or reduction operations.
\end{enumerate}

Any double-buffered kernel, responding to \Cref{eq:gemm-runtime}, meets condition 1 if the workload is communication-bound ($T_{comm} > T_{comp}$).
In this regime, reducing $T_{comm}$ results in a reduction of the total runtime $T$.
For any kernel in which $T_{comm}$ is an additive term in $T$, e.g. the reduction in FusedConcatLinear \gls{gemm}, condition 1 is always met and the speedup depends only on the fraction of $T_{comm}$ over $T$.

While we do not develop a full end-to-end network evaluation in this work, prior studies consistently show that GEMM-based operators dominate the runtime of many modern \gls{ml} workloads.
\cite{karami2024nongemmbench} reports that GEMM-based operators account for approximately 42.8\%-96.6\% of total inference latency across 17 widely used models spanning multiple tasks, while \cite{dice2021} shows that GEMM kernels account for 66.2\%-91.5\% of transformer inference runtime on \glspl{cpu}, depending on sequence length and thread count.
We therefore expect tangible improvements in end-to-end inference time, even though we focus improvements on GEMM runtime alone.

\section{Related Work}
\label{sec:related-work}


Numerous works have explored the design of multicast-capable on-chip networks, which can be classified into path-based and tree-based approaches.
Path-based methods deliver packets to destinations sequentially, resulting in higher latency but simpler designs \cite{lu2006, ebrahimi2010, ouyang2023, deng2025torrent}.
Conversely, tree-based methods replicate packets at intermediate routers, reducing latency at the expense of increased complexity.

This added complexity largely stems from the use cases and design goals targeted by prior works, which primarily focus on the one-to-many and many-to-one traffic patterns characteristic of cache-coherency protocols \cite{jerger2008, abad2009, krishna2011, krishna2014}, e.g. deriving from invalidation and acknowledgment messages.
To support such scenarios, prior architectures have prioritized flexibility, often at the cost of performance and scalability, through complex mechanisms handling arbitrary multicast patterns \cite{rodrigo2008, wang2009, wang2011, samman2008nocarc, hu2011, krishna2011, ma2012, zhong2014, samman2012, konstantinou2020}.
For example, several works adopt tag-based encodings to represent arbitrary destination sets, requiring additional tree setup \cite{jerger2008, samman2008date, ouyang2021, doe2020} or network partitioning \cite{rodrigo2008, wang2009} steps which lead to increased latency, while others employ highly flexible yet non-scalable destination encodings \cite{shen2017}.
Other works aim to balance link utilization \cite{wang2009, abad2009, krishna2011, samman2012}, a key factor in mitigating congestion caused by dense, irregular multicast traffic in cache-coherent shared-memory systems.
These approaches are, however, susceptible to deadlock, prompting the development of sophisticated deadlock avoidance or recovery mechanisms \cite{malumbres1996, samman2011, jerger2008, samman2012}.

In contrast to prior works focused on cache-coherent shared-memory systems, few have targeted \gls{ml} accelerators and their characteristic coarse-grained (or bursted), software-managed data transfers.
Recent studies have proposed designs tailored to multicast patterns in \gls{ml} workloads \cite{ouyang2021, ouyang2023}, but these remain specialized for narrow, fixed-function accelerator templates rather than general-purpose programmable systems.
One exception is \cite{colagrande2025}, which targets \gls{xbar}-based interconnects.
However, such interconnects offer limited scalability for large-scale, tile-based \gls{ml} accelerators \cite{fischer2025}.
From an industrial perspective, multicast support has also appeared in several commercial accelerator chips, such as Meta's \textit{MTIA} \cite{metamtia2023}, SambaNova's \textit{SN40L} \cite{sambanova2024} and Tenstorrent's \textit{Blackhole} \cite{tenstorrentblackhole2024} accelerators.
However, these solutions are proprietary and closed-source; the exact mechanisms, scalability characteristics, and performance implications are not publicly documented or quantitatively evaluated.
To the best of our knowledge, our work is the first to demonstrate scalable and open-source end-to-end hardware support for collective operations in programmable \gls{ml} accelerators.


On the other hand, conventional wisdom suggests that on-chip hardware support for many-to-one traffic should be avoided \cite{krishna2011} due to prohibitive area costs.
As a result, only a few studies have examined support for many-to-one collective operations, either focusing solely on ``gather'' primitives \cite{heisswolf2013, tiwari2020} or on combined multicast-reduction operations \cite{krishna2011, ma2012}, where reduction is limited to the aggregation of short acknowledgment messages.
To the best of our knowledge, our work is also the first to demonstrate high-throughput arithmetic reduction operations, traditionally deemed too costly for on-chip implementation, enabled by the \gls{dca} paradigm.

Finally, none of the previous works evaluate hardware-based collective primitives against optimized software baselines, as such comparisons can not be made for irregular, dynamic hardware-generated traffic.
The sole exception is \cite{colagrande2025}, which targets a different topology.
Other studies compare only software-based implementations \cite{barnett1995, barnett1996, matienzo2013}.
In contrast, we present a detailed comparison of hardware-accelerated and optimized software implementations, complemented by modeling efforts and an analysis on the relationship between the two approaches.


\section{Conclusion}

In this work, we introduced a lightweight collective-capable \gls{noc} tailored for large-scale next-generation \gls{ml} accelerators, extending the \gls{soa} FlooNoC architecture with hardware support for barrier synchronization, multicast and reduction operations.
At its core, the proposed \acrfull{dca} paradigm enables high-throughput in-network reduction operations with only \ResultAreaOverheadTot\% router area overhead.
Compared to highly-optimized software baselines, our hardware-accelerated multicast and reduction primitives achieve \ResultGeomeanHwMcastTwoDSpeedup$\times$ and \ResultGeomeanHwReductionTwoDSpeedup$\times$ geomean speedups, respectively, on transfers between \SI{1}{} and \SI{32}{\kibi\byte} of data.
Our evaluation is complemented by extensive modeling efforts, providing insights into the relationship between software and hardware collective implementations.
Finally, by keeping communication off the critical path of \gls{gemm} workloads, we estimate performance gains up to \ResultMaxGemmHwMcastSpeedup$\times$ and energy savings up to \ResultSummaMaxEnergySaving$\times$, demonstrating the benefits of in-network computation for scalable \gls{ml} acceleration.


\ifdefined\blindreview
Hidden for double blind review.
\else
\fi

\bibliography{paper}
\bibliographystyle{mlsys2026}


\appendix

\section{Hardware Terminology}
\label{sec:hw-terminology}

\subsection{AXI (Advanced eXtensible Interface)}

AXI~\cite{axi} is an on-chip communication protocol defined by Arm as part of the AMBA specification.
It organizes transactions into five independent channels: read address (AR), read data (R), write address (AW), write data (W), and write response (B), each using valid/ready handshaking.
AXI supports \emph{burst} transactions: a single address transaction (AR or AW) can be followed by multiple data transfers, each called a \emph{beat}, allowing large payloads to be streamed without repeating address-phase overhead.
AXI also supports outstanding transactions, i.e.\ multiple in-flight requests before prior responses are received, enabling high link utilization.
The \texttt{AWUSER} field is a user-defined sideband signal in the write address channel; this work uses it to extend the AXI protocol with collective communication support.

\subsection{RTL and Gate-Level Simulation}

\gls{rtl} simulation evaluates a hardware design expressed as registers connected by combinational logic, resolved cycle by cycle.
It enables cycle-accurate functional verification and performance measurement.
In this work, \gls{rtl} simulation is performed using \textsc{QuestaSim 2023.4}~\cite{questasim}.
Gate-level simulation operates on the synthesized netlist (the design expressed as standard library cells, i.e. logic gates and flip-flops), and can incorporate propagation delays from the physical layout.
Gate-level simulation is used here to extract realistic switching-activity data for power estimation with \textsc{PrimeTime 2022.03}~\cite{primetime}.

\subsection{Physical Design and EDA Flow}

Physical design maps the synthesized netlist to chip locations (placement) and connects them with metal wires (routing), producing a layout from which parasitic RC values are extracted for timing and power evaluation.
In this work, synthesis and place-and-route are performed with \textsc{Fusion Compiler 2024.09}~\cite{fusioncompiler}, a Synopsys EDA tool that combines logic synthesis and physical implementation in a unified flow.
\textsc{QuestaSim}~\cite{questasim} is a simulation tool by Siemens EDA supporting \gls{rtl} and gate-level simulation of \gls{rtl} designs.
\textsc{PrimeTime}~\cite{primetime} is a Synopsys tool for static timing analysis and power estimation from switching-activity data.

\section{Generalization to 2D Collectives}

\subsection{Multicast}
\label{sec:appendix-2d-mcast}

\begin{figure*}[t]
    \centering
    \includegraphics[width=0.56\textwidth]{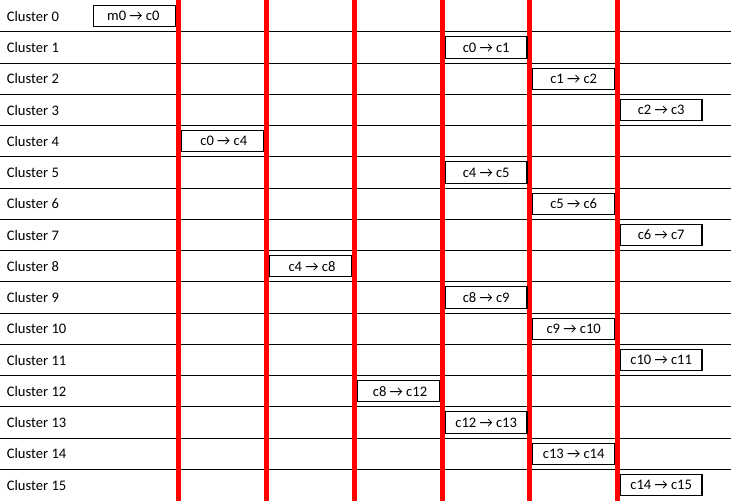}
    \caption{2D naive sequential multicast.}
    \label{fig:naive-mcast-2d}
\end{figure*}

\begin{figure*}[t]
    \centering
    \includegraphics[width=0.56\textwidth]{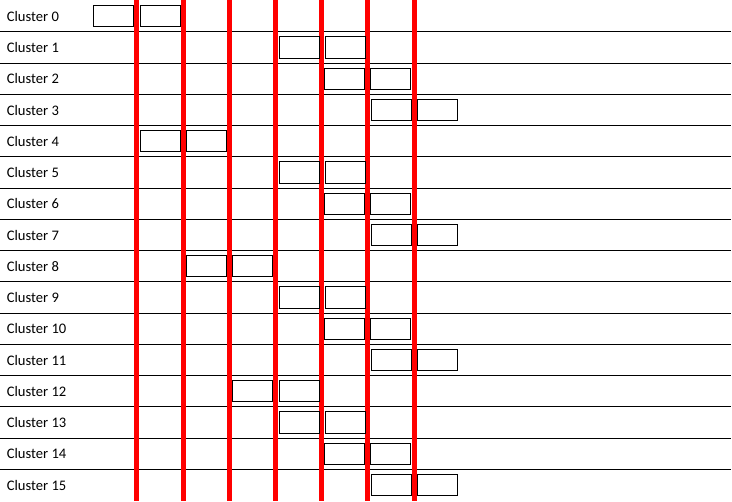}
    \caption{2D pipelined sequential multicast.}
    \label{fig:seq-mcast-2d}
\end{figure*}

\begin{figure*}[t]
    \centering
    \includegraphics[width=0.56\textwidth]{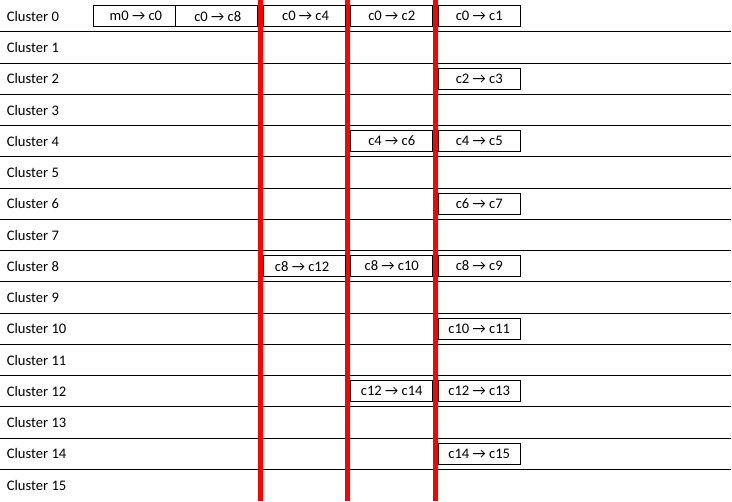}
    \caption{2D tree multicast.}
    \label{fig:tree-mcast-2d}
\end{figure*}

Formulas for the runtime of a 2D multicast transfer, corresponding to the \lstinline{naive}, \lstinline{seq} and \lstinline{tree} implementations depicted in \Crefrange{fig:naive-mcast-2d}{fig:tree-mcast-2d}:
\begin{equation}
    \label{eq:naive-mcast-2d}
    T_{naive} = \sum_{i=1}^{c+r-1} (\alpha_i + \beta n + \delta) - \delta
\end{equation}

\begin{equation}
    \label{eq:seq-mcast-2d}
    T_{seq} = \sum_{i=1}^{k + c + r - 2} (\alpha_i + \frac{n}{k} \beta + \delta) - \delta
\end{equation}

\begin{equation}
    \label{eq:tree-mcast-2d}
    T_{tree} = \sum_{i}^{\log_2 (c r)} (\alpha_i + n \beta + \delta) - 2 \delta
\end{equation}

The same formula for the \lstinline{hw} implementation:
\begin{equation}
    \label{eq:hw-mcast-2d}
    T_{hw} = \alpha + (n + c + r - 2) \beta
\end{equation}

\subsection{Reduction}
\label{sec:appendix-2d-reduction}

\begin{figure*}[t]
    \centering
    \includegraphics[width=\textwidth]{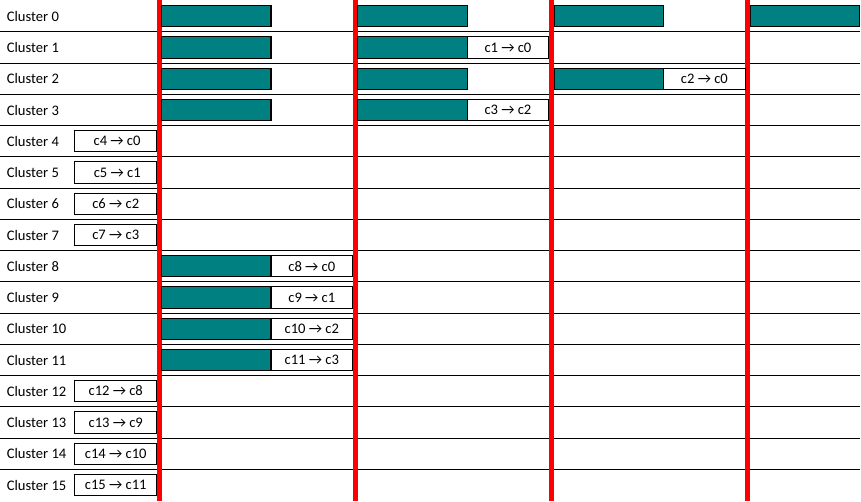}
    \caption{2D naive tree reduction.}
    \label{fig:naive-tree-reduction-2d}
\end{figure*}

\begin{figure*}[t]
    \centering
    \includegraphics[width=\textwidth]{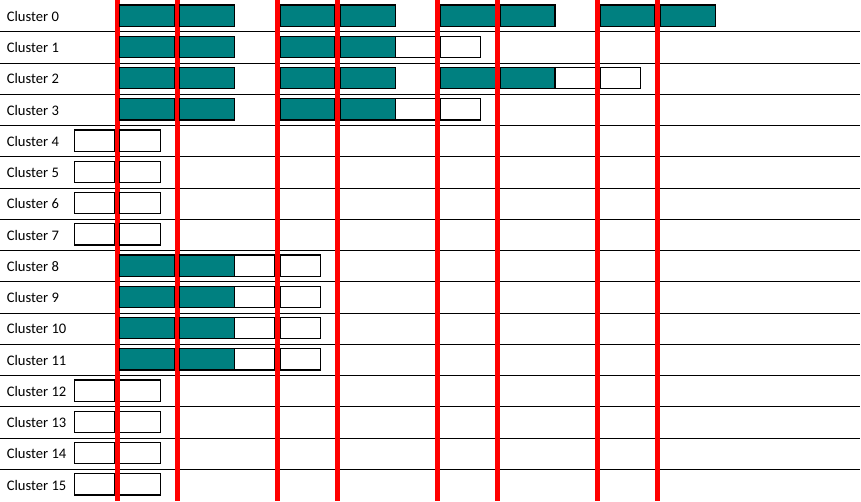}
    \caption{2D double-buffered tree reduction.}
    \label{fig:tree-reduction-2d}
\end{figure*}

\begin{figure*}[t]
    \centering
    \includegraphics[width=\textwidth]{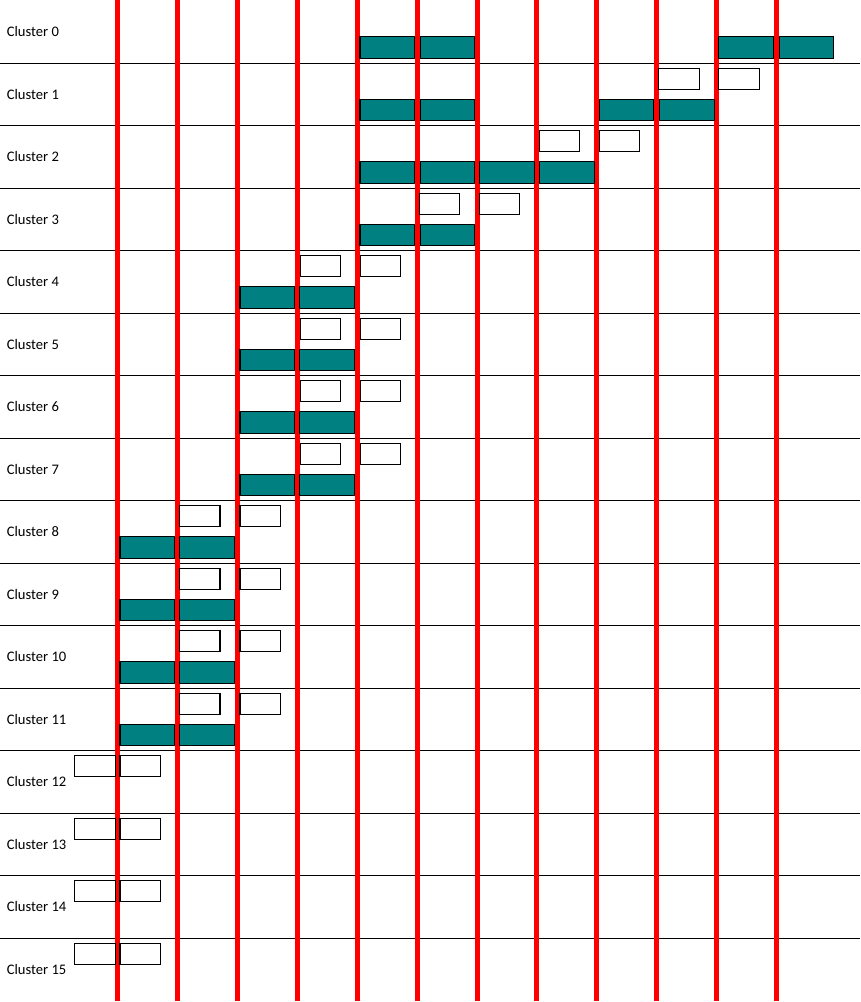}
    \caption{2D pipelined sequential reduction.}
    \label{fig:seq-reduction-2d}
\end{figure*}

Formulas for the runtime of a 2D reduction, corresponding to the \lstinline{tree} and \lstinline{seq} implementations depicted in \Crefrange{fig:tree-reduction-2d}{fig:seq-reduction-2d}:

\begin{equation}
    \label{eq:tree-reduction-2d}
    \begin{split}
        T_{tree} & = \left\{ t_m + \delta + (k - 1)\left[\max(t_m, t_c) + \delta\right] + \right. \\
                 & \left. + t_c \right\} * (\log_2 c + \log_2 r)
    \end{split}
\end{equation}

\begin{equation}
    \label{eq:seq-reduction-2d}
    \begin{split}
        T_{seq} & = t_m + 2(c - 2)\max(t_m, t_c) + (k-1) t_c + \\
                & + \max(t_m, t_c) + 2(r - 2)\max(t_m, t_c) + \\
                & + k t_c + (2(c - 2) + 2(r - 2) + 2k) \delta
    \end{split}
\end{equation}


\end{document}